\newif\ifmydraft

\mydraftfalse  

\ifmydraft
\documentclass[aps,prb,draft,superscriptaddress,showpacs,floatfix]{revtex4-1}
\usepackage{amsbsy}
\usepackage{amsmath}
\usepackage{amssymb}
\usepackage{amsfonts}

\else
\documentclass{ws-rv961x669.modified}
\usepackage{ws-rv-van}     
\fi
\makeindex

\begin{document}

\ifmydraft
\title{Wavefunctionology:  The Special Structure of Certain Fractional Quantum Hall Wavefunctions}
   \newcommand{\refcite}{\cite}
   \newcommand{\tbl}{\caption}
\else
\chapter[Wavefunctionology]{Wavefunctionology:  The Special Structure of Certain Fractional Quantum Hall Wavefunctions 
\label{ra_ctrueh1} }
\fi

\author[S.~H.~Simon]{Steven H. Simon}

\address{Rudolf Peierls Centre for Theoretical Physics\\
 	Clarendon Laboratory, Parks Road \\
Oxford, OX1 3PU, United Kingom \\
steven.simon@physics.ox.ac.uk}

\begin{abstract}

Certain fractional quantum Hall wavefunctions ---  particularly including the Laughlin, Moore-Read, and Read-Rezayi wavefunctions ---  have special structure that makes them amenable to analysis using an exeptionally wide range of techniques including conformal field theory (CFT), thin cylinder or torus limit, study of symmetric polynomials and Jack polynomials, and so-called ``special" parent Hamiltonians.   This review discusses these techniques as well as explaining to what degree some other quantum Hall wavefunctions share this special structure.  Along the way we will explore the physics of quantum Hall edges, entanglement spectra, quasiparticles,  nonabelian braiding statistics, and Hall viscosity, among other topics.  As compared to a number of other recent reviews, most of this review is written so as to {\it not} rely on results from conformal field theory --- although a short discussion of a few key relations to CFT are included near the end.

 \end{abstract}


\newpage
 
\tableofcontents
 
\ifmydraft
\maketitle
\else
\body
\fi

\section{Introduction}\index{Introdution}

The study of fractional quantum Hall effect (FQHE) is unusual within condensed matter  in that we are often able to write down explicit wavefunctions that are extremely accurate representations of what we believe is realized in experiment.   Starting with the seminal work of Laughlin\cite{laughlintheory}, the field has often progressed by constructing zero-parameter ``variational" trial wavefunctions.   While these wavefunctions are usually not exact descriptions of any experiment, they are often extremely precise approximations.  More importantly these trial wavefunctions are believed to describe the same universal physics as is observed in experiment.  A modern approach to this relation is to state that the experimental system and the trial wavefunction may be deformed into each other without closing the excitation gap --- thus meaning that they are in the same (topological) phase of matter. 

The study of these wavefunctions ---  their mathematical structure, their excitations, their  Hamiltonians, their entanglement properties,  the braiding statistics of their quasiparticles, and so forth  --- has become a huge endeavor.  
Even though this endeavor has been pursued for almost forty years, even now new ideas are being uncovered about the structure and properties of these wavefunctions.   The purpose of this chapter is to review a number of the important advances in understanding that have occurred in the last few decades. 


Since the seminal work of Moore and Read\cite{moore1991nonabelions}, conformal field theory (CFT) has played a central role in the understanding of FQHE.   Despite this fact, many of the 
important results and ideas, although closely related to CFT, can be exposited without detailed discussion of CFT.  Particularly since the application of CFT to FQHE has been reviewed elsewhere recently\cite{nayakreview,HanssonRMP}, the intent of this review  is instead to describes some of the interesting new material with minimal explicit use of CFT.      Hence this review should complementary to other already existing reviews.   (That said, in the final section of this chapter (section \ref{sec:CFT}), I will turn to briefly discuss a few connections to CFT that are crucial enough, or are modern enough, to warrent additional discussion here.)

Much of this review will focus on the Laughlin states\cite{laughlintheory}, the Moore-Read state\cite{Moore89}, and the Read-Rezayi series\cite{read1999beyond}.  These wavefunctions are special for several reasons.   Perhaps most importantly, they arise as the ground state of simple  special parent Hamiltonians such that the ground state as well as the quasihole excitations and edge excitations, have exactly zero interaction energy (See section \ref{sec:parent}).  This property greatly simplifies their understanding.    Further, not only the ground state, but also all of the zero interaction energy states (edge excitations and quasiholes) may be explicitly represented as CFT correlators (See section \ref{sec:CFT}).  

There will be also be some discussion of other  wavefunctions, such as the Gaffnian\cite{Gaffnian} (and also the Haffnian\cite{dmitrygreenthesis})  which also have very simple special parent Hamiltonians.    As we will discuss further these wavefunctions do not correspond to gapped phases of matter, but are nonetheless intseresting.   Many other wavefunctions sharing at least some of the nice properties of these wavefunctions can also be constructed using Jack polynomials\cite{BernevigHaldane1} which we discuss in section \ref{sec:Jack}. 

We will also briefly discuss a few aspects of the experimentally important hierarchy\cite{haldane1983fractional,halperin1983theory} (or composite fermion\cite{jain2007composite})  phases of matter   although this topic is somewhat de-emphasized given the recent review on hierarchy physics\cite{HanssonRMP}. 

We will begin  our discussion with some quantum Hall basics in section \ref{sec:basics}.  We introduce the structure of the lowest Landau level in the sphere, cylinder, and disk geometries.   We then describe the properties we should demand of our many-body wavefunctions in section \ref{sub:many}.

In section \ref{sec:parent} we discuss parent and ``special" Hamiltonians.  We begin by discussing in section \ref{sub:laughlin} how the Laughlin wavefunction can be constructed as the ground state of a special Hamiltonian.   We discuss special parent Hamiltonians in section \ref{sub:RRandMR} for the Read-Rezayi series with special attention on the Moore-Read state.  In section \ref{sub:otherparent} the extent to which these ideas can be extended to other wavefunctions is discussed. 

Once we've described special parent Hamiltonians for quantum Hall states it is easy to discuss the space of edge excitations as being wavefunctions that do not increase the interaction energy.  In section \ref{sec:edge} we explore this approach to edge physics in general. We start with the Laughlin edge in section \ref{sub:laughlinedge}, and explain how this corresponds to a one dimensional bose theory.    We also discuss how the inner product between edge state excitations can be described using an effective bosonic edge theory in section \ref{sub:innerprod}.    In section \ref{sub:hierarchy} we (briefly) study the edge of hierarchy states.  Then in section \ref{sub:rrandmooreedge} we discuss the more complicated edge of the Moore-Read state (with brief discussion of Read-Rezayi edges). 

In section \ref{sec:thin} we introduce the idea of a thin-cylinder limit, which simplifies a lot of the intuition about certain quantum Hall states --- such as finding the ground state degeneracy on a torus.   We show in section \ref{sub:edgestatecounting} how the thin limit allows us to determine the edge spectrum for even complicated wavefunctions corresponding to special parent Hamiltonians.   In section \ref{sec:Jack} we explain how the thin-limit is closely related to the Jack polynomial approach to writing wavefunctions for bulk systems.    

In section \ref{sec:entanglement} we invoke ideas from quantum information theory and explain the idea of an entanglement spectrum and explain how that is related to the edge spectrum. 

In section \ref{sec:quasiholes} we turn to describe wavefunctions involving localized  quasiholes, beginning with the Laughlin state.  First we discuss how the localized quasiholes can be thought of as a superposition of edge excitations.  We then move on to the Moore-Read and Read-Rezayi states.   The existence of many degenerate quasihole wavefunctions in these cases is a signature of their nonabelian statistics, which we describe briefly in section \ref{sub:nonabelian}.   We then describe how nonabelian braiding statistics can be calculated in section \ref{subsub:braiding} as a combination of monodromy and Berry matrix, and explain the importance of working with an orthonormal holomorphic basis of wavefunctions in section \ref{subsub:holo}.  

In section \ref{sec:HallVisc} we discuss Hall viscosity and explain both its relation to Berry phase and also its relation to more conventional fluid dynamical response functions.   

In section \ref{sec:CFT} we turn to give a few comments on conformal field theory and its relationship to the states discussed here.   We emphasize the idea of bulk-edge correspondence, and we discuss the importance of conformal blocks and the key issue of their orthonormality.    In section \ref{sub:specialCFT} we discuss what is special about the CFTs that produce well known quantum Hall states. 

Finally, in section \ref{sec:conclusions} we offer some brief concluding remarks.

\section{Some FQHE basics} 
\label{sec:basics}

\subsection{Single Particle Physics} 

We begin with some obligatory basics of FQHE, mainly to establish notation and language   (see Refs.~\refcite{prangebook,chakrabortybook,jain2007composite} for more details of basics). We consider a two dimensional gas of particles in a high magentic field $B$ perpendicular to the plane of the sample.  These particles, with some effective mass $m$, may be fermions (like electrons, as in actual FQHE experiments to date) or they may be bosons (which may someday\footnote{There is one experimental preprint claiming to have produced FQHE with cold atoms\cite{Gemelke}.  While this is an extremely interesting work, it has still not been confirmed or followed up.   Very recent work\cite{JonathanSimon} has convincingly observed a Laughlin state of two bosons of light interacting via coupling to Rydberg atoms.}
be engineered to produce FQHE, see chapter by Nigel Cooper in this volume or Ref.~\refcite{CooperReview}).  Note that it is a common abuse of nomenclature to call the particles contituting our system ``electrons", independent of whether they are bosons or fermions.  

The particles interact with each other via some interaction potential (to be specified in more detail below) and we will often confine the particles to a droplet by appling some potential $U({\bf r})$.   For simplicity we will ignore any additional degrees of freedom which may exist in certain experiments, such a particle spin (which we can assume is polarized) or valley (as in graphene, see chapter by Dean, Kim, Li, and Young in this volume), or layer index (See Ref.~\refcite{EisensteinReview}) .  Our Hamiltonian (our ``theory of everything") can then be written as 
$$
H = \left( \sum_{i}  \frac{[{\bf p}_i - e {\bf A}({\bf r}_i)]^2}{2m^*} \right)+ \left( \sum_{i}  U({\bf r_i}) \right) + H_{ \mbox{\scriptsize interaction}}
$$
The first term in brackets, the kinetic energy, we assume is the largest scale in the problem.  The single particle spectrum of this kinetic energy term breaks into highly degenerate Landau levels with energies $E_n = \hbar \omega_c (n+1/2)$ with $\omega_c = e B/m^*$ being the cyclotron frequency given the effective mass $m^*$.  We assume that a single Landau level is partially filled, which, without loss of generalitye  we can assume is the lowest Landau level (i.e., $n=0$)\footnote{A partially filled higher Landau level can be mapped to a partially filled lowest Landau level at the price of modifying the interaction and confinement terms, so long as we can neglect inter-Landau-level transitions.\cite{prangebook,simonrezayicooperpseudo}.}.    

If we work in symmetric gauge in a planar geometry\cite{prangebook,chakrabortybook,jain2007composite}, in the absence of a confining potential $U$, the single particle eigenstates (of the kinetic term of the Hamiltonian) in the lowest Landau level can be written as
\begin{equation}
 \varphi_m(z) \propto  \,\, z^m \,\,\,  \mu(z,\bar z) \label{eq:varphim}
\end{equation}
with 
\begin{equation}
 \mu(z,\bar z) = e^{-|z|^2/4}
\end{equation}
where here $z=x + i y $ is the complex representation of the particle position in the plane, and we have set the magnetic length $\ell = \sqrt{\hbar/(e B)}$ to unity.   Here, $m$ is the angular momentum of the eigenstate around the origin (the complex phase wrapping $m$ times) and the shape of the wavefunction is roughly an annulus having radius $\sqrt{2m}$ and thickness roughly unity.   Thus, the radius is linked to the angular momentum of the eigenstate.  

Often we consider fractional quantum Hall systems on surfaces other than a plane.   On the sphere\cite{haldane1983fractional}, one places a monopole with magentic flux $N_\phi$ at the center of the sphere, and the lowest Landau level then has exactly $N_\phi+1$ orbitals.  Using\cite{ReadRezayi1996} stereographic projection from the sphere to the plane, $z = 2 R \tan(\theta/2)e^{i \phi}$ with $R$ the radius of the sphere, the wavefunction of these orbitals in an appropriately chosen gauge can be written in form very similar to that of Eq.~\ref{eq:varphim} except that on the sphere
\begin{equation}
   \mu(z,\bar z) =    \left(1 + \frac{|z|^2}{4 R^2}\right)^{-1-N_\phi/2}_.
\end{equation}
The orbitals $z^m$ again have the shape of an annulus of thickness approximately one, at constant lattitude around the sphere, going from the north pole ($m=0$) to the south pole ($m=N_\phi$).   

On the cylinder geometry of circumference $L$, we use complex coordinate $w=x+iy$ with the real direction going around the circumference and the imaginary direction going along the length of the cylinder and we define $z= e^{2 \pi i w/L}$.   The single particle orbitals are of the form\cite{MilovanovicRead} of Eq.~\ref{eq:varphim} except on the cylinder
\begin{equation}
 \mu(z,\bar z) = e^{-|y|^2/2}
\end{equation}
The $z^m$ orbitals form rings around the circumference with thickness approximately unity, and whose distance along the cylinder is indexed by $m$.   Note that since the cylinder may be infinitely long in both directions, $m$ can be any integer (including negative).    However, we will often consider no orbitals with $m<0$ to be occupied in order to make more simple analogy with the sphere and plane wavefunctions.   One can, of course connect up the ends of the cylinder to form a torus as well\cite{Haldane1985periodic}.

These three geometries (plane, sphere, cylinder) are all useful.  Since the interesting part of the wavefunction are the $z^m$ factors, it is fairly easy to translate physics from one geometry to another.  We will typically work in whichever geometry is simplest for elucidating the interesting physics.  

\subsection{Many Body Physics} 
\label{sub:many}

A many-body wavefunction $\Psi$ for $N$ particles is generally written as 
$$
 \Psi(z_1, \ldots, z_N) = \Phi(z_1, \ldots, z_N) \,\, \prod_{i=1}^N \mu(z_i,\bar z_i)
$$
where $\Phi$ is a polynomial in the $z's$  which is fully symmetric for bosons or fully antisymmetric for fermions.  We will typically only write the $\Phi$ part of the wavefunction for simplicity.   

If the wavefuncton represents an angular momentum eigenstate (as, for example, a ground state would if it is confined in a rotationally invariant potential) then the polynomial $\Phi$ should also be homogeneous in degree, since each power of $z$ contributes one quantum of angular momentum to the wavefunction. 

The ratio of the number of particles $N$ to the number of available single-particle orbitals $N_{orb}$ in the Landau level is known  as the filling fraction 
$$
\nu= \lim_{N  \rightarrow \infty}  \frac{N}{N_{orb_.}} 
$$  
To determine the filling fraction of a many-body wavefunction, we need to compare the number of particles to the number of orbitals which are at least partially filled.  Indicating orbitals as $z^n$, let the lowest value of $n$ (the lowest angular momentum orbital) which is at least partially occupied be $n=0$.  On the cylinder this assumption fixes the location of the one of the edges of the quantum Hall droplet, whereas on the sphere or plane this simply states that the quantum Hall droplet covers the north pole or origin of the plane. 
Then we look for the highest power of $z$ in the wavefunction (which we will call ``maxpower") to find the the farthest away orbital which corresponds to the largest radius of the droplet.  The total number of orbitals which are at least partially filled is then $N_{orb} = ({\rm maxpower}+1)$.  Thus the filling fraction is $\nu=N/({\rm{maxpower}}+1)$.    

On the sphere we have the more detailed relationship for a quantum Hall state covering the entire sphere
\begin{equation}
\label{eq:shiftdef}
 N_{\phi} = \frac{1}{\nu} N - {\cal S}
\end{equation}
where $\cal S$ is an order one number known as the {\it shift} of the wavefunction\cite{WenZeeShift} and the flux at the center of the sphere is $N_\phi={\rm maxpower}=N_{orb} -1$ for the Lowest landau level.   This definition of the shift is most cleanly defined on a sphere where one can precisly determine the flux in the sphere and count the number of electrons necessary to fill the sphere (leaving no quasielectrons or quasiholes) for any particular quantum  Hall state. For the disk or cylinder, one defines $N_\phi$ as $N_\phi = N_{orb}  -1$ where $N_{orb}$ is the number of orbitals at least partially occupied.  

Note that even for the integer quantum Hall effect, the shift is an interesting quantity (See for example the discussion in Ref.~\refcite{jain2007composite}).   Filling the lowest Landau level on the sphere requires $N = N_{\phi}+1$ electrons, whereas filling the first excited Landau level requires $N=N_{\phi}+3$.   The shift turns out to be twice the mean orbital spin of an electron in the given state\cite{read2009nonabelian,read2011hall} --- which depends, for example, on which Landau levels we are considering (see also the discussion in section \ref{sec:CFT} below). 

One can multiply a wavefunction by an overall Jastrow factors to produce a new wavefunction
\begin{eqnarray} \label{eq:multJas}
 \Phi^{\mbox{\scriptsize new}} &=&  \Phi^{\mbox{\scriptsize old}}  \prod_{i<j} (z_i - z_j)^p \\
\nu_{\mbox{\scriptsize new}}&=&\frac{1}{\nu_{\mbox{\scriptsize old}}^{-1} +p
}  
\end{eqnarray} 
Indeed, this transformation is the root of the entire composite fermion approach\cite{jain2007composite}. (See also the chapter by Jainendra Jain in this volume).   Note that if the old wavefunction describes bosons (fermions), the new wavefunction will describe bosons (fermions) for even $p$ and will describe fermions (bosons) for odd $p$.

There is an immense freedom  (i.e., an enormous Hilbert space) for writing down wavefunctions for partially filled Landau levels.   For fermions, when there are more orbitals available than there are fermions to fill them, we have an enormous degeneracy of states ($N_{orb}$ choose $N$)  corresponding to deciding where these fermions should go.  For bosons since we can multiply occupy single orbitals, a degeneracy ($N_{orb} -1 + N$ choose $N$) exists even when there are more bosons than orbitals.   This enormous (exponentially large in $N$ for fixed $\nu$) degeneracy is broken by the interactions between the particles and, under appropriate circumstances, a unique FQHE ground state is formed. 

We thus turn to consider the interaction term of the Hamiltonian.  In real FQHE experiments, the interaction term involves a two-body interaction
$$
  H_{\mbox{\scriptsize  interaction}} = \sum_{i < j} V({\bf r}_i - {\bf r}_j)
$$
with $V$ being, for example, the Coulomb interaction.  From a theoretical standpoint, however, we will often find it useful to think of more general many-body interactions such as a three-body interaction
$$
\sum_{i < j < k} V({\bf r}_i - {\bf r}_j, {\bf r_i} - {\bf r_k}).
$$
or even interactions involving a larger number of particles.   Note that, starting with a two-body interaction,  integrating out inter-Landau-level transitions will generate many body interactions\cite{SimonRezayiPert,SodemannMacDonald,PetersonNayak}.

\section{FQHE Wavefunctions and Special Parent Hamiltonians}
\label{sec:parent}

\subsection{Laughlin}
\label{sub:laughlin}

Laughlin's genius was to simply guess the right wavefunction that very accurately described the experimentally observed $\nu=1/p$ fractional quantum Hall effect\cite{laughlintheory}:  
\begin{equation}
 \Phi_{\mbox{\scriptsize Laughlin}}^{\nu=1/p} = \prod_{i < j} (z_i-z_j)^p~~~_.  
 \label{eq:Laugh} 
\end{equation}
Note that for odd $p$ this is an antisymmetric wavefunction appropriate for fermions, whereas for even $p$ it is symmetric and appropriate for bosons.    Just to check that this does indeed describe filling fraction $1/p$ we note that the highest power of any $z$ occurring is $z^{p(N-1)}$.   Thus, for example, on a sphere, our wavefunction would cover the sphere perfectly if $N_{\phi} = \mbox{maxpower}  = p (N-1)$.  In the large system limit, then we would have a ratio $\nu=N/N_{orb} = N/(N_\phi + 1) \rightarrow 1/p$.   

Now although this wavefunction is exceedingly accurate for certain real physical systems (say, electrons in high mobility GaAs quantum wells at filling fraction $\nu=1/3$), for theoretical work it is useful to consider a situation for which this wavefunction will be the {\it exact} ground state.  This will enable us to make precise statements, after which we can think about whether (or to what degree) our statements carry over to the physical experiments.   For this reason it is useful to think about so-called {\sl parent Hamiltonians} which give the desired wavefunctions as their unique exact ground states.   A Hamiltonian that is ``parent" also has the property that the interaction energy is non-negative definite and is exactly zero in the ground state.  A further property, which we will call a ``special" parent Hamiltonian\cite{Read2009}, is that all low energy edge excitations and quasihole excitations also have zero interaction energy.   Much of this review is focused on properties of quantum Hall states that are generated by special parent Hamiltonians. 

A simple example is the special parent Hamiltonian\cite{haldane1983fractional,trugman,Pokrovsky} for the bosonic $\nu=1/2$ Laughlin state (the $p=2$ case of Eq.~\ref{eq:Laugh}) which is obtained by using the interaction 
\begin{equation}
\label{eq:deltainteraction}
 H_{\mbox{\scriptsize interaction}} =  V_0 \sum_{i< j}   \delta({\bf r}_i - {\bf r}_j)
\end{equation}
with $\delta$ being a two-dimensional delta-function and $V_0 > 0$ an interaction energy scale.  It is easy to see that the Laughlin $\nu=1/2$ wavefunction makes the energy of this interaction exactly zero --- since in the wavefunction there is zero amplitude for two particles coming to the same position.  One should be cautious, however, that the energy would also be zero for any wavefunction that is a fully symmetric polynomial times the Laughlin $\nu=1/2$ wavefunction  (it would have to be a symmetric polynomial to maintain the symmetry of the wavefunction) since the wavefunction would still vanish when any two particles come to the same position.   However, multiplying the Laughlin wavefunction by a polynomial will increase the radius of the droplet, since, as discussed above (just below Eq.~\ref{eq:varphim}), increasing the degree of the polynomial increases the radius of the wavefunction.  Thus, if we include a weak radially symmetric confining potential $U(|{\bf r}|)$, the lowest energy wavefunction, and hence the unique exact ground state, will be the Laughlin wavefunction itself$\,$\footnote{As we will discuss in section \ref{sec:edge} below, if the confining potential is quadratic $U({\bf r}) \propto |{\bf r}|^2$ there is substantial further simplification and the confinement need not be weak.}$^,$\footnote{It is worth commenting that even for the Laughlin wavefunction with this very simple special parent Hamiltonian, there is no {\it rigorous} proof that the bulk is gapped in the thermodynamic limit, and this reasonable assumption is justified mainly by numerical work. \label{foot:gapped}}.    Multiplying by a polynomial generates low energy edge excitations as we will discuss in section \ref{sec:edge}.

It is obvious that a delta function interaction is ineffective for fermions since two fermions cannot come to the same position anyway.  However, we can still write a special parent Hamiltonian for the fermionic $\nu=1/3$ case by using a Laplacian of a delta function\cite{haldane1983fractional,trugman,Pokrovsky}.  In general, special parent Hamiltonians can be written for any Laughlin state  $\nu=1/p$ for arbitrary $p$ by considering high enough derivatives of delta functions.   

Perhaps a more transparent way to describe such a parent Hamiltonian interactions is via the idea of pseudopotentials\cite{haldane1983fractional,prangebook} often written as $V_m$.  We can write {\it any} 2-body interaction within a single Landau level as  
$$
 H_{\mbox{\scriptsize interaction}} =  \sum_m \sum_{i< j}  V_m  P_{ij}^m
$$
where $P_{ij}^m$ is a projection operator that projects the wavefunction such that particles $i$ and $j$ have relative angular momentum $m$, i.e., when\cite{simonrezayicooperpseudo}
$$
  \psi_{rel}  \sim (z_i - z_j)^m 
$$
So, for example if we have an interaction with $V_0 > 0$ this gives positive energy to any wavefunction that does not vanish when two particles come to the same position (this is precisely the delta function interaction of Eq.~\ref{eq:deltainteraction}).  If we have $V_1 > 0$ this gives energy to a wavefunction that vanishes as $(z_i - z_j)$ when two particles come to the same position (this is the Laplacian of a delta function), and so forth.   So for example, if we want to write a special parent Hamiltonian for the Laughlin state of bosons at $\nu=1/4$ we simply set $V_0 > 0$ and $V_2 > 0$, which gives positive energy unless the wavefunction vanishes as four or more powers when two particles approach each other.   By exchange symmetry, a bosonic (fermionic) wavefunction can only vanish as an even (odd) number of powers as two particles approach each other, so that for bosonic (fermionic) systems only the even (odd) pseudopotentials matter.

\subsection{Read-Rezayi and Special Case of Moore-Read}
\label{sub:RRandMR}

The idea of psuedopotentials have been generalized to interactions involving more than two particles\cite{simonrezayicooperpseudo,PapicClustering}.   A simple application of this idea\cite{SimonRezayiCooperHamiltonian} is to define the  Hamiltonian 
\begin{equation}
H_k^r =  \sum_{i_1 < i_2 < \ldots i_{k+1}} P_{i_1, i_2, \ldots i_{k+1}}^r
\label{eq:Hkrdef}
\end{equation}
where $P^r_{i_1, \ldots i_{k+1}}$ is a projection operator that projects the cluster of $k+1$ particles to have relative angular momentum greater or equal to $r$.   Thus, $H_k^r$ gives positive energy unless the wavefunction vanishes at least as fast as $r$ powers when $k+1$ particles come to the same point.   For example, the Laughlin $\nu=1/p$ Hamiltonian is $H_1^p$.   

A crucial example of this construction is given by the Hamiltonian $H^2_k$ which is the special parent Hamiltonian for the bosonic $\mathbb{Z}_k$ Read-Rezayi wavefunction\cite{read1999beyond}. The explicit real-space form of this Hamiltonian is given by
\begin{equation}
 H_k^2 =   V_0^{(k)} \!\!\!\!\sum_{i_1 < i_2 < \ldots i_{k+1}}  \delta({\bf r}_{i_1} - {\bf r}_{i_2}) \delta({\bf r}_{i_1} - {\bf r}_{i_3})  \ldots \delta({\bf r}_{i_1} - {\bf r}_{i_{k+1}})  \label{eq:kplus1bodydelta}
\end{equation}
with $V_0^{(k)} > 0$.   This interaction allows $k$ particles to come to the same point, but gives an energy penalty when $k+1$ come to the same point.  Equivalently $H_k^2$ forbids a cluster of $k+1$ particles from having zero relative angular momentum, but allows relative angular momentum of $2$ hence the superscript on the $H_k^2$ (Note that a cluster of bosons cannot\cite{SimonRezayiCooperHamiltonian} have relative angular momentum of 1). 

The ground state of the $H_k^2$ Hamiltonian can be written as the following generalization of the Laughlin state (where the number of particles $N$ is a multiple of $k$), known as the ${\mathbb{Z}}_k$ Read-Rezayi wavefunction\cite{read1999beyond,cappelli2001parafermion} 
\begin{eqnarray} \nonumber
 \Phi_{\mbox{\scriptsize Read-Rezayi}}^{\nu=k/2; \mbox{ \scriptsize bosons}} &=& {\mathbb S}\left[  \left(\prod_{0<i<j \leq N/k}(z_i - z_j)^2 \right)\left(\prod_{N/k<i<j \leq 2N/k}(z_i - z_j)^2 \right) \ldots  \right. \\
 & & \left.  \ldots \left(\prod_{N(k-1)/k<i<j \leq N}(z_i - z_j)^2 \right)  \right]   \label{eq:RRwavefunction}
\end{eqnarray}
Here, ${\mathbb S}$ means symmetrize over all reordering of the particle numbering.   So the scheme here is to divide all the particles into $k$ equal groups and write a Laughlin $\nu=1/2$ wavefunction for each group, then symmetrize over all different ways you might have chosen these groups.  It is easy to see why this is a zero energy wavefunction of the $(k+1)$-body delta-function:  for a cluster of $k$ particles all coming to the same position, the wavefunction need not vanish, since we can put one particle in each of the $k$ groups.  However, when the $k+1^{st}$ particle comes to that point, then we must have (at least) two particles in (at least) one group and the wavefunction then must vanish.   Indeed, while perhaps not entirely obvious, it can be shown that, analogous to the Laughlin case, this is the most dense wavefunction (the lowest degree polynomial) with this property and is hence the ground state. 
Note that the $k=1$ case of the Read-Rezayi series is the Laughlin wavefunction and the $k=2$ case is known as the Moore-Read state\cite{moore1991nonabelions} which we will discuss in more depth in section \ref{subsub:mooreread}.

We can attach Jastrow factors to the the Read-Rezayi series (plugging Eq.~\ref{eq:RRwavefunction} in Eq.~\ref{eq:multJas}) to obtain wavefunctions with filling fractions
\begin{equation}
 \nu = \frac{k}{2 + k p}
\label{eq:RRnu}
\end{equation}
with $p \geq 0$ even for bosonic wavefunctions, and $p>0$ odd for fermionic wavefunctions.   Note that $k=1$ corresponds to the Laughlin series. 

It is also straightforward to write special parent Hamiltonians for the Read-Rezayi wavefunctions with Jastrow factors attached.   In order to enforce $p$ Jastrow factors in our wavefunction, we must include a two body term ($H_1^p$) in our Hamiltonian which is precisely the special parent Hamiltonian for Laughlin $\nu=1/p$.    Once we have these $p$ Jastrow factors, bringing $k+1$ particles to the same point will incur $pk(k+1)/2$ powers of $z$.   Thus we can include another term in the Hamiltonian that forbids $k+1$ particles from coming together with relative angular momentum  less than or equal to $pk(k+1)/2$. Such a Hamiltonian is exactly $H^{pk(k+1)/2 + 2}_k$.   Thus the total Hamiltonian is $H^{pk(k+1))/2}_k + H_1^p$.  In the case where we only want to include a single Jastrow factor for a fermionic wavefunction ($p=1$), we do not need to include the term $H_1^1$ since it is trivial for fermions. 

By counting powers of $z$ and looking for the highest power that occurs (see the discussion near Eq.~\ref{eq:shiftdef}), we can find the precise relationship between flux and particle number for the Read-Rezayi wavefunction with $p$ Jastrow factors attached:
\begin{equation}
\label{eq:Nphiequation}
  N_\phi = \left(\frac{2}{k} + p\right) N - (p+2)
\end{equation}
which clearly gives the filling fraction $\nu$ in Eq.~\ref{eq:RRnu} in the large system size limit, and gives the shift ${\cal S} = p+2$.

\subsubsection{Special case of Moore-Read}
\label{subsub:mooreread}

The $k=2$ case of the Read-Rezayi series deserves special attention.  This is known as the Moore-Read wavefunction\cite{moore1991nonabelions}, and it can be rewritten in a different form:
\begin{equation}
\Phi_{\mbox{\scriptsize Moore-Read}}^{\nu=1;  \mbox{\scriptsize bosons}} = {\rm Pf} ({\bf M}) \prod_{i < j} (z_i - z_j)
\label{eq:PfEq} 
\end{equation}
where here $\bf M$  is the antisymmetric matrix with components 
\begin{equation}
\label{eq:PfM}
M_{ij} = \frac{1}{z_i  - z_j}
\end{equation}
and zeros along the diagonal.  The notation 
${\rm Pf} ({\bf M})$ symbolizes a Pfaffian, which, for an $N$ by $N$ matrix (with $N$ even) is given by
$$
 {\rm Pf}({\bf M})  = \frac{1}{2^N (N!)} \,\, {\mathbb A} \left[M_{12}M_{34}  \ldots M_{(N-1),N} \right]
$$
where here ${\mathbb A}$ is the antisymmetrized sum over all permutations of indices\footnote{ Note the useful identity   
$
[{\rm Pf}({\bf M})]^2 = {\rm det}({\bf M})_.
$
}.
It is interesting to note that all BCS superconducting wavefunctions are essentially Pfaffians\cite{degennesbook}.  For Cooper pairing of spinless fermions, the BCS wavefunction is precisely a Pfaffian where $M_{ij} = g({\bf r}_i - {\bf r}_j)$ with $g({\bf r})$ the (necessarily antisymmetric)  wavefunction of a single pair.  Thus, the Moore-Read wavefunction is essentially a spinless superconductor.

\subsection{Other Parent Cluster Hamiltonians and their Wavefunctions}
\label{sub:otherparent}

Parent Hamiltonians have been written down for a number of other wavefunctions, including, in particular, cases including wavefunctions with multiple spin states\cite{halperin1983theory,ardonne2001nonabelian,HaldaneRezayiWavefunction,
SchoutensRead1,SchoutensRead2,BLOK,SpinSingletGaffnian,ReadRezayi1996}.  Generalizatons of parent Hamiltonians to lattice fractional quantum Hall effect\footnote{Particles hopping on a lattice which exhibit fractional quantum Hall effect are known as Fractional Chern Insulators\cite{BergholtzReview,PARAMESWARAN}.} have been pursued by a number of authors\cite{KapitMullerLaughlin,KapitMullerNonAbelian,
Nielsen1,Nielsen2,Greiter1,Greiter2,BergholtzModel}.

Returning to spinless particles in the continuum, an obvious extension\cite{SimonRezayiCooperHamiltonian,PapicClustering} of the  parent Hamiltonians for Laughlin and Read-Rezayi wavefunctions is to consider Hamiltonians $H_k^r$  that give positive energy to a cluster of $k+1$ particles with relative angular momentum less than some number $r$ (i.e., forces the wavefunction to vanish as at least $r$ powers when $k+1$ particles approach the same point) which we mght expect to give a wavefunction at filling fraction $\nu=k/r$ for bosons.  An interesting case is that of $k=2,r=3$, which for bosons gives a $\nu=2/3$ wavefunction known as the Gaffnian wavefunction\cite{Gaffnian}.  The fermionic version at $\nu=2/5$ (the wavefunction being obtained by attaching a single Jastrow factor to the bosonic $\nu=2/3$ wavefunction) is obtained using $k=2,r=6$.  The bosonic Gaffnian wavefunction has extremely high overlap with the Jain $\nu=2/3$ bosonic quantum Hall state\cite{Gaffnian}; and similarly the fermionic Gaffnian at $\nu=2/5$ has extremely high overlap with the Jain $\nu=2/5$ state.   However, where the composite fermion wavefunctions correspond to gapped phases of matter, the Gaffnian, which is related to the nonunitary $M(3,5)$ conformal field theory,  is actually gapless\cite{Gaffnian,JoliGaffnian,Read2009,RegnaultScreening,ByungminKang,Galois}, having zero energy neutral excitations in the thermodynamic limit.  Indeed, it has been argued that any wavefunction related to a nonunitary conformal field theory must similarly be gapless\cite{Read2009}.  A rough argument as to why nonunitarity in a CFT implies gaplessness in the bulk is given by the bulk-edge correspondence which will be discussed in section \ref{sec:CFT} below\cite{Read2009}.   One possibile scenario is that the Gaffnian represents a critical point between two gapped quantum Hall phases --- although this remains a conjecture.

Another cluster wavefunction worth discussing is known as the Haffnian\cite{dmitrygreenthesis}, corresponding to $k=2,r=4$ which is a $\nu=1/2$ wavefunction for bosons, with a fermionic version being $k=2, r=7$ at $\nu=1/3$ .   The Haffnian (related to a unitary, but {\it non-rational} conformal field theory\cite{Senechal97}) is also believed to be gapless and is even more poorly behaved than the Gaffnian --- having an extensive ground state degeneracy on a torus in the thermodynamic limit\cite{PapicClustering}. 

Although it may seem that one should be able to generate many new wavefunctions with such clustering Hamiltonians, the situation is not as simple as it seems\cite{SimonRezayiCooperHamiltonian}.   For simplicity let us consider the case of boson wavefunctions.    For the cases we have discussed thus far (Laughlin, Moore-Read, Read-Rezayi, Gaffnian, Haffnian), we have used the Hamiltonian $H_k^r$ which forbids the wavefunction from vanishing as fewer than $r$ powers when $k+1$ particles approach each other.  In each of these cases we get a ground state wavefunction where two special properties hold:  (a) the wavefunction does not vanish when $k$ particles approach each other and (b) the wavefunction vanishes as exactly $r$ powers when $k+1$ particles approach each other.   As long as these two properties hold we will obtain a wavefunction with filling fraction\cite{SimonRezayiCooperHamiltonian}
\begin{equation}
\label{eq:nukr}
 \nu(k,r) = k/r_.
\end{equation}
However, the ground state of $H_k^r$ need not satisfy these two properties for general $k$ and $r$.  The ground state can vanish when fewer than $k+1$ particles come together, or it can vanish as more than $r$ powers when $k+1$ particles come together\cite{SimonRezayiCooperHamiltonian}.   A prime example of this behavior is the Hamiltonian $H_2^5$ for bosons, which requires that the wavefunction vanish as (at least) $r=5$ powers when $k+1=3$ particles come to the same point.  However, the ground state\cite{SimonRezayiCooperHamiltonian} of this Hamiltonian is actually the $\nu=1/2$ Laughlin state, which vanishes when only two particles come to the same point, and vanishes as 6 powers when three come to the same point.  It turns out to be possible\cite{Jackson} to analytically construct a special parent Hamiltonian for a wavefunction that vanishes as exactly $r=5$ powers as $k+1=3$ particles come to the same point (related to the nonunitary $M(3,7)$ wavefunction), however it is somewhat more complicated.  

In all of the above cases (Laughlin, Moore-Read, Read-Rezayi, Gaffnian, Haffnian, $M(3,7)$), we have specified that the wavefunction must vanish as $r$ powers when $k+1$ particles come to the same point.   It has been proposed that specifying these vanishing properties (a so-called ``pattern of zeros") can be used to define a wavefunction\cite{WenWangPattern1,WenWangPattern2,WenWangPattern3}.
While some results can be derived from thinking in this language, there is also a limitation.   Unfortunately, the simple statement that a wavefunction vanishes as $r$ powers as $k+1$ particles come to the same point is not generally sufficient information to fully define a wavefunction.   To clarify this, let us consider a situation where we have only $k+1$ particles. 
When there is more than one (translationally invariant) polynomial of $r^{th}$ degree in $k+1$ variables, the vanishing properties alone do not define the polynomial. \cite{SimonRezayiCooperHamiltonian,Liptrap,Jackson,PapicClustering}.   This situation occurs for $k=2, r=6$ and $k=2, r \geq 8$  and all $k>2, r>3$.  Instead, in these cases there is a multidimensional space of $(k+1)$-particle wavefunctions which vanish as $r$ powers as all $k+1$ particles come together.    This implies that the pattern of zeros approach needs to be supplemented with additional information in order to uniquely define a particular wavefunction.

In these cases where there is a multidimensional space of polynomials available for a $(k+1)$-particle wavefunction, the projection operator in the Hamiltonian of Eq.~\ref{eq:Hkrdef} can be refined and written as a projection onto a subspace of available wavefunctions.  We thus choose to give positive energy to some subspace of this while leaving the remaining subspace (possibly a single $(k+1)$-particle wavefunction) at zero energy\cite{SimonS3,Jackson}.   In this way it is still possible to write parent Hamiltonians for wavefunctions which vanish proportional to a particular polynomial function as some number ($k+1$) particles come to the same point.   In these more complicated cases, special parent Hamiltonians have successfully been constructed 
for 
$k=2,r=6$ ($\nu=1/3$ for bosons as dictated by Eq.~\ref{eq:nukr}, which are related to the $N=1$ superconformal field theories with generic\footnote{The cases corresponding to the $N=1$ superconformal minimal series are more complicated.} central charge\cite{Jackson,SimonN1}).  It was also shown\cite{Jackson} that one can construct special parent Hamiltonians corresponding to $M(3,p)$ conformal field theories.  These correspond to $k=2, r=p-2$ and thus to $\nu=2/(p-2)$ for bosons.  Note that except for the simple Moore-Read state (which is $p=4$) these are all nonunitary conformal field theories and are therefore gapless (see section \ref{sec:CFT} below). 

Other examples of parent Hamiltonians have been explored where the ground state is also picked out by specifying a particular polynomial describing the wavefunction when $k+1$ particles come together.  These include the case\cite{SimonS3} of $k=3,r=4$  ($\nu=3/4$ for bosons, as dictated by Eq.~\ref{eq:nukr}) related to the $S_3$ conformal field theories, and also the minimal models of $N=1$ superconformal field theories $k=2, r=6$  ($\nu=1/3$ for bosons).  However, in these cases it has so far not been possible to fully determine how the zero energy excitations match up with the predictions of the CFT.    Nonetheless, while it appears possible to construct parent Hamiltonians (but perhaps not special parent Hamiltonians) for wavefunctions which correspond to unitary conformal field theories (and therefore do not run afoul of the rule\cite{Read2009} that nonunitary (or nonrational) CFTs must give gapless wavefunctions) it remains an open question whether these Hamiltonians correspond to gapped phases of matter.

An obvious question is whether one can construct special parent Hamiltonians for the experimentally prominent hierarchy\cite{haldane1983fractional,halperin1983theory}, or composite fermion\cite{jain2007composite}, series of fractional quantum Hall states (See chapter by Jain in this volume).  While no exact parent Hamiltonians have been constructed, one approach for studying the composite fermion state at $\nu=2/3$ for bosons (or correspondingly 2/5 for fermions) uses a delta-function (or derivative of delta function) interaction within two degenerate Landau levels\cite{JainEarly,RezayiMacdonald,Seidel25}.  While the resulting wavefunction appears to have many of the properties desired, it is not fully projected to the lowest Landau level, which is problematic.  (Generalizing this approach to three degenerate Landau levels\cite{WenParton,SeidelParton} gives a parent Hamiltonian for a so-called {\it parton} wavefunction\cite{WenParton,JainEarlier} related to the $SU(2)_2$ conformal field theory.)  In general, determing simple parent Hamiltonians for hierarchy or composite fermion wavefunctions remains an important open problem\cite{Fremling}.   It should be noted, however that given any wavefunction corresponding to a gapped phase of matter, one can in principle generate a local parent Hamiltonian, at least numerically\cite{Greiter1,Greiter2,Chertkov,Ranard}.  However, the resulting Hamiltonian generically can involve terms with many particle interactions, and will not usually be simple.     Further, it is unclear if such a Hamiltonian will have the ``special" property that the low energy quasiholes and edge excitations have exactly zero interaction energy.   

\section{Edge Excitations}
\label{sec:edge}

We now turn to consider the elementary excitations of the FQHE states described by special parent Hamiltonians.   Since the bulk of the system is assumed to be gapped$^{\ref{foot:gapped}}$, when the confining potential is taken into account, we expect that any low energy excitations will involve excitations of the edge only.   The study of quantum Hall edges is a venerable subject\cite{Halperin81,WenChiral}, and we will only cover some aspects, referring to the literature for a more complete discussion\cite{KaneFisher,Chang}.

A physical picture of the edge excitation is given by classical physics.  A FQH droplet is assumed to be held in position by some confining potential, which we assume to be rotationally  symmetric so that angular momentum remains a good quantum number.   This can be thought of as some radial electric field acting on our charged particles and pushing them inwards.   On the other hand, we also have a magnetic field perpendicular to the plane of the sample.  From basic classical physics, since we have a crossed electric and magentic field we then obtain a drift velocity for charged particles, pushing them perpendicular to the edge (perpendicular to both electric and magnetic fields).   So if a small bump is created at some point along the edge, it will travel around the edge of the droplet at some drift velocity.  This is a simple way to picture the edge excitations. 

Things simplify quite a bit if we choose to use a parabolic confinement for our quantum Hall droplet on a disk
$$
 U({\bf r}) = \gamma|{\bf r}|^2.
$$
In this case, despite this perturbation to the kinetic term of the Hamiltonian, the single particle eigenstates still take the simple form of Eq.~\ref{eq:varphim} (although the magnetic length is altered by the confining potential).  With the confining potential, the energy of each single particle eigenstate in the lowest Landau level is now given by 
\begin{equation}
\label{eq:Emlin}
 E_m^{LLL} =  \alpha m + \mbox{constant}
\end{equation}
where the prefactor $\alpha$ is set by the strength $\gamma$ of the confinement\footnote{In the limit of large magentic field, we have $\alpha=2\gamma $ as expected.}.    The intuition here is that the eigenstate of angular momemtum $m$ is at radius $r = \sqrt{2 m}$ and thus has an energy $\gamma r^2 \propto m$.  See the discussion near Eq.~\ref{eq:varphim}.   Note that in the cylinder geometry one instead obtains an edge excitation of the form of Eq.~\ref{eq:Emlin} for a linearly sloped edge potential $U({\bf r}) \sim y$. 

The ground state of the special parent Hamiltonian with confinement is the (presumably unique) wavefunction that has zero interaction energy (i.e., satisfies the constraints imposed by the interaction) and also has the minimal total angular momemtum, and therefore has the minimal possible confinement energy consistent with zero interaction energy.     To look for low energy excitations,  we need to find a polynomial wavefunctions of slightly higher degree than the ground state, which still satisfies the zero interaction energy constraint imposed by the special parent Hamiltonian.   By being a slightly higher degree polynomial, the droplet will be of slighly higher radius, and slightly higher confinement energy.   For each added degree of the polynomial, the wavefunction increases its angular momentum by one unit, and the energy increases by one unit of the confinement energy $\alpha$. 

\subsection{Laughlin Edge}
\label{sub:laughlinedge}

For example, consider the special parent Hamiltonian for the Laughlin $\nu=1/p$ state for a system bosons (in the case of even $p$) or fermions (for odd $p$).  Here the special parent Hamiltonian interaction is such that we must find a (symmetric for bosons, antisymmetric for fermions) polynomial which vanishes as $p$ or more powers when two particles come to the same position, and we want the polynomial to be of minimal total degree.   Thus there should be a factor of $(z_i - z_j)^p$ for each pair of particles.  This uniquely picks out the Laughlin $\nu=1/p$ wavefunction as the ground state as discussed above.    To obtain low energy edge excitations, as discussed below Eq. \ref{eq:deltainteraction}, we multiply the Laughlin wavefunction by any symmetric polynomial.   The product still satisfies the constraint that the wavefunction vanishes as at least $p$ powers when two particle approach each other and maintains proper bosonic (fermionic) symmetry of the wavefunction for $p$ even (odd).   Thus enumerating the possible edge excitations of the Laughlin state amounts to enumerating all possible symmetric polynomials by which we can multiply the ground state wavefunction\cite{WenChiral,StoneSchur}.

There are many ways to achieve the goal of enumerating symmetric polynomials.  Perhaps the simplest method is to use the so-called power sum symmetric polynomials.   
\begin{equation}
\label{eq:pmdef}
 p_m = \sum_{i=1}^N  z_i^m
\end{equation}
which form the generators of the {\it ring} of symmetric polynomials\footnote{A ring is roughly a set of objects that we can add and multiply.    The statement that $p_m$ are generators means that sums and products (and multiplying by a constant) of these $p_m$'s form all possible symmetric polynomials.}.  (We will see other ways to enumerate all symmetric polynomials in section \ref{sec:Jack} below). 
We can then make a table  (Table \ref{tab:sympoly}) of the number of polynomials we can write at degree $d$, and hence the number of edge eigenstates we can write whose angular momentum is $d$ units more than that of the ground state. 

\begin{table}
\tbl{Symmetric polynomials listed by degree, and the energy of the eigenstate which results when they multiply the Laughlin ground state.
 }
{
\begin{tabular}{c|l|l}
degree $d$  & ~~~~~~~~~~~ symmetric polynomials \rule[15pt]{20pt}{0pt} &    $E - E_{\mbox{\scriptsize ground}}$ \\
\hline
1	& $~~p_1~~$ & $\phantom{1} \alpha$ \\
2    & $~~p_2, ~~~~ p_1 p_1 ~~~ $ &  $2 \alpha$ \\
3    & $~~p_3, ~~~~ p_2 p_1, ~~~~ p_1 p_1 p_1$ &  $3 \alpha$ \\
4    & $~~p_4, ~~~~ p_3 p_1, ~~~~ p_2 p_1 p_1,  ~~~~  p_2 p_2, ~~~~ p_1 p_1 p_1 p_1$~~~~ &  $4 \alpha$ 
\end{tabular}}
\label{tab:sympoly}
\end{table}

There is a unique ground state (the Laughlin state).  Examining Table \ref{tab:sympoly} we see there is a single eigenstate with energy $\alpha$ greater than the ground state which is given by multiplying the ground state by $p_1$.   The space of eigenstates with energy $2 \alpha$ greater than the ground state is spanned by two wavefunctions, one of which is $p_1^2$ times the grounds state and the other is $p_2$ times the ground state (thus the space is two dimensional).  For higher angular momentum, eigenstates with angular momentum $q$ units greater than the ground state will have energy $q \alpha$ greater than the ground state, and the number of such eigenstates with will be equal to the number of integer partitions of the integer $q$.  This then gives the well known series 1,2,3,5,7,11, $\ldots$, as shown in Table \ref{tab:sympoly}.  

The wavefunctions built from the products of $p_m$ operators multiplied by the Laughlin wavefunction provide a linearly independent set of states spanning the Hilbert space of excitations at a given angular momentum.   However, we are not guaranteed that these wavefunctions are orthogonal to each other (for example, $p_1^2$ times Laughlin and $p_2$ times Laughlin span the space of $q=2$ excitations, but are not strictly orthogonal). However, quite interestingly it turns out that in the thermodynamic limit (large enough droplet and small enough $q$), these wavefunctions do indeed become orthogonal\cite{WenEdgeReview,DubailReadRezayi,Fern2018A} as we will discuss further below in section \ref{sub:innerprod}.   

We now ask whether we can write an effective theory of the edge that does not require us to work with wavefunctions for all $N$ particles in the entire FQHE droplet --- after all, we are only concerned with (potentailly small) deformations of the edge not the entire system.   Indeed we can write such a theory, which is known as a chiral Luttinger liquid\cite{WenChiral,WenEdgeReview,KaneFisher} 
\begin{equation}
\label{eq:xLL}
H_{\chi LL} = \sum_{n > 0} (\alpha n) a^\dagger_n a_n^{\phantom{\dagger}}
\end{equation}
Here, the operator $a^\dagger_n$ ($a_n$) is a bosonic operator corresponding to the creation (annihilation) of the $n^{th}$ angular momentum mode of the edge.   Since they are bosonic operators they satisfy the usual commutations 
\begin{equation}
[a_n,a^\dagger_m] =  \delta_{nm.}
\label{eq:acomm}
\end{equation}
Here the modes have only positive angular momentum $n > 0$, indicating that modes only travel in one direction (i.e., the exciations are chiral).   If we think of the Laughlin ground state as being the ``vacuum" $|0\rangle$ of the edge (i.e., the edge with no excitations), we can then make a table of all excitations in order of their angular momentum (and hence of their energy) as shown in Table \ref{tab:cll}.

\vspace*{10pt}
\begin{table}
\tbl{Excitations of the chiral Luttinger liquid (Hamiltonian given in Eq.~\ref{eq:xLL}) and their corresponding energies.  The first column is the angular momentum compared to the ground state.}
{\begin{tabular}{c|l|l}
~~~~$L-L_{ground}$  & ~~~~~~~~~~~~~~~states~~~\rule[15pt]{20pt}{0pt}  &    Energy  \\
\hline 
1	& $~~a^\dagger_1|0
\rangle~~~~~$ & $\phantom{1} \alpha $ \rule[10pt]{15pt}{0pt}\\
2    & $~~a^\dagger_2|0
\rangle, ~~~ a^\dagger_1 a^\dagger_1|0
\rangle ~~~~ $ &  $2 \alpha $ \rule[10pt]{15pt}{0pt} \\
3    & $~~a^\dagger_3|0
\rangle, ~~~ a^\dagger_2 a^\dagger_1|0
\rangle , ~~~ a^\dagger_1 a^\dagger_1  a^\dagger_1|0
\rangle $ &  $3 \alpha $ \rule[10pt]{15pt}{0pt}\\
4    & $~~a^\dagger_4|0
\rangle,~~~ a^\dagger_3 a^\dagger_1|0
\rangle,~~~
 a^\dagger_2 a^\dagger_1  a^\dagger_1|0
\rangle,~~~~ a^\dagger_2 a^\dagger_2 |0
\rangle,~~~~ a^\dagger_1 a^\dagger_1 a^\dagger_1  a^\dagger_1|0
\rangle $ &  $4 \alpha $ \rule[10pt]{15pt}{0pt} \\
\end{tabular}
}
\label{tab:cll}
\end{table}

We immediately notice that Tables \ref{tab:sympoly}  and  \ref{tab:cll} are essentially identical.   With the mapping
$$
 p_m \leftrightarrow a^\dagger_m
$$
there is a precise correspondence between the two descriptions of the edge\cite{DubailReadRezayi,Fern2018A}.   As noted above, the $p_m$'s give orthogonal states only in the thermodynamic limit (large system and low energy of excitation), whereas the $a^\dagger_m$'s are orthogonal by definition.    We discuss this issue in more depth in section \ref{sub:innerprod}.

\subsubsection{Bosonized Edge}
\label{subsub:bosonized}

Although this material has been discussed many places previously\cite{KaneFisher,WenChiral,Chang,HanssonRMP}, it is worth discussing it again here for completeness.   The chiral Luttinger liquid Hamiltonian density for the $\nu=1/p$ Laughlin state can be written in terms of a bosonic field $\phi(x)$ as 
\begin{equation}
 {\cal H} =  \frac{v}{4 \pi \nu}(\partial_x  \phi)^2
 \label{eq:xLL2}
\end{equation}
where $v$ is the edge velocity, and the commutations of the field $\phi$ are given by
\begin{equation}\label{eq:phicom}
 [\phi(x), \phi(x')] = i \nu \pi {\rm{sign}}(x - x')_,
\end{equation}
and the charge density along the edge is given by $\rho(x) = \partial_x \phi/(2 \pi)$. 

To make contact with Eq.~\ref{eq:xLL} we consider the system on a disk as in Eq.~\ref{eq:xLL} and write $\phi(z)$ with $z$ the complex position around the edge.  In this case we can precisely write
\begin{equation}
\label{eq:phidef}
  \phi(z) =  \phi_0 -i  a_0 \ln z + i \sum_{n > 0} \frac{1}{\sqrt{n}} \left( a_n  z^{-n}
+  a_n^\dagger z^{n}\right) 
\end{equation}
where the $a^\dagger_n$ are the same bosonic mode creation operators as in Eq.~\ref{eq:xLL} and  where $a_0$ is the so-called zero-mode operator that counts the total charge in the system and $[\phi_0,a_0]=i$ defines the conjugate operator $\phi_0$ which changes the number of particles in the system.\cite{Senechal97,DubailReadRezayi,Fern2018B}. 

Note that in the language of this bosonic field, we can write a charge bump with total charge $\gamma$ at position $x$ along the edge as
$$
 V_\alpha = :e^{i (\gamma/\nu) \phi(x)}:
$$
with colons indicating normal ordering. In particular for $\gamma=1$ we obtain an electron creation operator.

\subsubsection{Further Laughlin Edge Considerations} 

Dimensional (RG) arguments tell us that the chiral Luttinger liquid Hamiltonian, Eq.~\ref{eq:xLL}  is the correct low energy long wavelength limit of the Laughlin edge\cite{KaneFisher,WenChiral}, independent of the Hamiltionian of the underlying problem (i.e., independent of the details of the interelectron interaction, although the value of $\alpha$ appearing in the Hamiltonian will depend on details)\footnote{\label{foot:lowenergy}The low energy long wavelength validity of the chiral Luttinger liquid Hamiltonian is true so long as ground state is in the Laughlin phase of matter, and so long as the edge does not ``reconstruct"\cite{WanReconstruction,Sabo2017,chamonwen1994,JohnsonEdgeReconstruction,MacDonaldEdgeReconstruction}, which is guaranteed when the inter-electron interaction is sufficiently close to the special parent Hamiltonian.  In practice there is a very wide range of edge structures and inter-electron interactions for which this remains correct.\cite{WanReconstruction}}.   At higher energy scales or at shorter wavelength the Hamiltonian Eq.~\ref{eq:xLL} (equivalently Eq.~\ref{eq:xLL2}) remains exact for the parent Hamiltonian, but for generic inter-electron interactions there will be subleading terms in the Hamiltonian which are less relevant in the renormalization group sense.  The study of the effect of such nonlinear terms in Luttinger liquids has been a topic of some interest outside of the quantum Hall field\cite{Imambekov}, and recently has attracted attention in the quantum Hall context as well\cite{Fern2018B,Lamacraft1,Lamacraft2,Bettelheim2006,Wiegmann2012,Abanov2005}.
There was some debate\cite{Bettelheim2006,Wiegmann2012} over whether, even without long range inter-electron interactions,  there could be long-range (so-called ``Benjamin-Ono") interaction terms in the effective edge Hamiltonian.   However, recent studies\cite{Fern2018B} support the view that, so long as the inter-electron interaction is not long ranged, additional terms in the effective edge Hamiltonian must also be short ranged.  Further, it was found that, for a droplet in a parabolic potential, there are a large number of additional constraints on the form of the edge Hamiltonian.  For example, writing down terms of the Hamiltonian in order of decreasing RG relevance, one might expect that the first correction to Eq.~\ref{eq:xLL2} would be a term of the form $(\partial \phi)^3$ (Indeed, this is a main correction that is studied in detail in Refs.~\refcite{Imambekov,Lamacraft1,Lamacraft2}).   However with parabolic confinement it was shown that the coefficient of this term must be exactly zero\cite{Fern2018B}.     It remains an outstanding question to determine whether constraints of this type are relevant for experimental systems in non-parabolic confinement.

\subsubsection{Edge State Inner Products}
\label{sub:innerprod}

As mentioned just before the start of section \ref{subsub:bosonized}, the edge state operators $a^\dagger_m$ correspond exactly to the polynomials $p_m$ with which we can multiply the ground state wavefunction --- so long as we are in the limit of large system size and low energy.  When we are not in this limit the corrections to this orthonogonality are given by an extremely interesting ``edge action" construction\cite{DubailReadRezayi,Fern2018A}.   Defining {\boldmath$\lambda$} to be a list of nonincreasing integers $\lambda_m$, we can write an arbitrary excited edge state as 
\begin{equation}
| {\mbox{\boldmath$\lambda$}}\rangle =   \prod_m  (a_m^\dagger)^{\lambda_m} | 0 \rangle  \label{eq:edgea}
\end{equation}
corresponding to the two dimensional wavefunction
\begin{equation}
\Psi_{\mbox{\boldmath$\lambda$}} =   \left(\prod_{m} p_m^{\lambda_m} \right) \Psi^{\nu}_{Laughlin}  \label{eq:edgepsi}
\end{equation}
with  (now normalized) power sum polymomials (compare Eq.~\ref{eq:pmdef})
\begin{equation}
  p_m = \frac{1}{\sqrt{\nu}} \sum_{i=1}^N \left(\frac{z_i}{R}\right)^m_.
\end{equation}
with $R$ the radius of the droplet.  Note that Eq.~\ref{eq:edgea} is a statement about a state on a one-dimensional edge, whereas Eq.~\ref{eq:edgepsi} describes the full two dimensional wavefunction of the droplet.   The inner product between two such edge states is generally nontrivial
\begin{equation}
\label{eq:edgeinner}
 \langle \Psi_{\mbox{\boldmath$\lambda$}} | 
 \Psi_{\mbox{\boldmath$\lambda'$}} \rangle =  \langle   
 \mbox{\boldmath$\lambda'$} | e^{-S} | \mbox{\boldmath$
 \lambda$} \rangle
\end{equation}
where $S$ has local interaction terms only, and which operates on the edge space.  This operator in the case of a disk geometry can be expanded as
$$
 S = -\frac{\sqrt{\nu}}{6 N} (\partial_z \phi)^3 + \ldots
$$
where each successive term is smaller by $1/\sqrt{N}$.   In the large system size limit, one needs only keep the leading term $e^{-S} \approx 1$ so that states which are orthogonal in the edge space are also orthogonal wavefunctions in the bulk.  

It is of course necessary that there is some mapping between states described as edge excitations (Eq.~\ref{eq:edgea}) and states described as bulk wavefunctions (Eq.~\ref{eq:edgepsi}).  What is not obvious is that the form of the mapping should be the exponential of local interaction terms on the edge space and that the operators should be increasingly small as the system gets larger.  This form can be justified via intuition from conformal field theory and matrix product states\cite{DubailReadRezayi}.  Although this justification does not constitute a rigorous proof that Eq.~\ref{eq:edgeinner} is correct, it has also been checked numerically to fairly high precision\cite{DubailReadRezayi,Fern2018A}.

\subsection{From Laughlin Edge To Hierarchy}
\label{sub:hierarchy}

The Haldane-Halperin hierarchy of quantum Hall states\cite{haldane1983fractional,halperin1983theory,halperin1984statistics} has recently been reviewed elsewhere\cite{HanssonRMP} so the  discussion here will remain brief. 
As mentioned above, the hierarchy states (which are topologically equivalent to the Jain hierarchy\cite{jain2007composite,read1990excitation} and should therefore have similar edge structure), do not have simple parent Hamiltonians.  Nonetheless, even without such Hamiltonians that we can solve exactly,  we still know the low energy edge structure of these states of matter.   Generally the edge will consist of multiple bosonic modes, which we describe by bosonic fields $\phi^a$ for $a=1, \ldots, n$ for some $n$, having commutation relations (analogous to Eq.~\ref{eq:phicom})
$$
 [\phi^a(x), \phi^b(x')] = i \pi K^{-1}_{ab} {\rm sign}(x - x')
$$
for some symmetric matrix of integers ${\bf K}$. 

In addition to the $\bf K$ matrix, a hierarchy quantum Hall states is defined by a vector $\bf t$ of integers which describe the physical charge of the 
respective edge modes via $\rho_a(x) = t_a \partial_x \phi^a/(2 \pi) $.  The filling fraction of the bulk is given by $\nu={\bf t}^T {\bf K}^{-1} {\bf t}$.   For example, in the traditional Haldane-Halperin hierarchy construction $t_1 =1$ and $t_a =0$ for $a > 1$, and 
$$
 K_{ab} = p_a \delta_{ab} - \delta_{a,b-1} - \delta_{a, b+1}
$$
with (for electrons making up the state) $p_1$ odd and all other $p_a$'s are even and can be positive or negative but not zero.  This generates filling fractions of the continued fraction form
$$
 \nu = {\bf t}^T {\bf K}^{-1} {\bf t} = \frac{1}{p_1 - \frac{1}{p_2 - \frac{1}{\ldots - \frac{1}{p_n}}}}
$$ 
Note that one is free to make a basis transformation  $
{\bf K} \rightarrow {\bf W}^T {\bf K} {\bf W}$ and ${\bf t} \rightarrow {\bf W}^T {\bf t}$ using any integer valued matrix $\bf W$ of unit determinant\cite{read1990excitation,WenEdgeReview}.   

Another useful example is that of the Jain series $\nu=p/(2p+1)$ which has $p$-dimensional $\bf K$ matrix and $\bf t$ vector given by  
\begin{equation}
\label{eq:JainK}
 K_{ab} = 2 + \delta_{ab}; ~~~~~~~~ t_a = 1
 \end{equation}
which can be converted to the hierarchy form via $W_{ab} = \delta_{ab} - \delta_{a-1,b}$.

The Hamiltonian density for the edge is then given (analogous to Eq.~\ref{eq:xLL2}) by
\begin{equation}
{\cal H} = \frac{1}{4 \pi} V_{ab} (\partial_x \phi^a) (\partial_x \phi^b)
\label{eq:xLLmultimode}
\end{equation}
where $V_{ab}$ is a nonuniversal interaction matrix which determines the $n$ edge mode velocities. 

Note that in addition to $\bf K$ and $\bf t$, another vector, conventionally called $\bf s$, representing the spin of the respective edge modes, is required to fully define the topological order of the quantum Hall states\cite{WenZeeShift,WenEdgeReview}.   This vector also determines the shift of the quantum Hall state (See Eq.~\ref{eq:shiftdef}) via 
$${\cal S} = \frac{2}{\nu} {\bf t}^T {\bf K^{-1}} \bf s.
$$  For example, for the Jain states described by Eq.~\ref{eq:JainK} the spins are given by $s_a = a+1/2$.  As with the $\bf t$ vector, the $\bf s$ vector transforms as ${\bf s} \rightarrow {\bf W}^T \bf s$ under basis transformation. 

To a large extent, one can describe the low energy edge excitations via multiple chiral Luttinger liquid Hamiltonians of the form of Eq.~\ref{eq:xLL} albeit with different velocities for each mode which can be in either the positive or negative directions.
One should also be cautious because there generally will also be low energy excitations associated with moving charge from one edge mode to another.  Generically at higher energy (or for smaller system sizes) there will be subleading terms of the Hamiltonian Eq.~\ref{eq:xLLmultimode} which are irrelevant in the renormalization group sense and are unimportant at low energy and large length scale.   The scaling dimension of subleading terms simply counts the total number of derivatives, so for example, $(\partial_x \phi)^3$ and $(\partial_x^2 \phi)^2$ are subleading compared to the leading term $(\partial_x \phi)^2$.

\subsection{Read-Rezayi and (mostly) Moore-Read Edges}
\label{sub:rrandmooreedge}

Let us now consider the case of a Read-Rezayi quantum Hall droplet.  As an example, let us consider the bosonic ${\mathbb{Z}}_k$ Read-Rezayi state at filling fraction $\nu=k/2$.  As mentioned above, the special parent Hamiltonian forces the wavefunction to vanish when $k+1$ particles come to the same position.   The resulting wavefunction is of the form of Eq.~\ref{eq:RRwavefunction}. 

As described above at the beginning of section \ref{sec:edge}, we consider a droplet with parabolic confinement, so that the edge spectrum is necessarily linear in angular momentum.   We must now figure out how many eigenstates there are at each angular momentum.   In other words we want to know how many linearly independent homogeneous symmetric polynomials there are at a given degree which satisfy the required condition that they must vanish as $k+1$ particles come to the same position\cite{Feigin}. 

As in the Laughlin case above, it is certainly possible to mulitply a Read-Rezayi ground state wavefunction by a symmetric polynomial of $q^{th}$ degree in order to generate a new wavefunction with $q$ units of angular momentum more than the ground state that still satisfies the required vanishing condition.   Doing this generates a bosonic edge mode excitation entirely analogous the above described Laughin case.   

However, there are other ways to generate acceptable wavefunctions.   In Eq. \ref{eq:RRwavefunction} we could multiply the first group by a symmetric polynomial of $q_1^{th}$ degree in the variables $z_1 \ldots z_{N/k}$, then we multiply the second group by another symmetric polynomial of $q_2^{th}$ degree in the variables $z_{N/k+1}, \ldots z_{2N/k}$ and so forth.  Within each group the polynomial still vanishes whenver two particles come to the same point, and since there are $k$ groups, it is necessarily the case that the wavefunction will vanish whenever $k+1$ particles come to the same point (there must be at least two particles within some group).    This strategy gives us vastly more possible ways to generate new polynomials satisfying our clustering rules.   Unfortunately, not all such polynomials will be linearly independent from each other --- and determining the dimension of the space of eigenstates turns out to be a rather tricky task\cite{ArdonneKedemStone,ReadRRCounting}. 

For the case of the Moore-Read (the $k=2$ Read-Rezayi) state, the counting of edge modes was achieved by Milovanovic and Read\cite{MilovanovicRead} (along with the counting of edge modes of the 331 state and the Haldane-Rezayi state, which both have multiple spin states).   The result of their work is suprisingly simple: assuming an even number of particles in the system, the Moore-Read edge is described by a bosonic mode (as in Laughlin) along with an additional Majorana edge mode.  We can thus write an effective theory for the Moore-Read edge as 
\begin{equation}\vspace*{5pt}
 H_{MR-edge} = \sum_{n > 0} (\alpha n) a^\dagger_n a_n^{\phantom{\dagger}} + \sum_{m \geq 0;\, m \, \mbox{\scriptsize odd}} \alpha (m/2)  \, \psi^\dagger_{m/2} \psi^{\phantom{\dagger}}_{m/2} \label{eq:HMR} 
\end{equation}
where $a$ is bosonic (satisfying Eq.~\ref{eq:acomm}) as in the Laughlin case, but $\psi$ is fermionic, satisfying 
$$
\{ \psi^\dagger_{m/2}, \psi^{\phantom{\dagger}}_{n/2} \} = \delta_{nm}
$$ 
with the added constraint that the parity of the number of fermions excited on the edge must match the parity of the number of electrons in the system.   We can build up a similar table of the possible excitations at small angular momenta which is shown in Table  \ref{tab:mr}.
\begin{table}
\tbl{Excitations of the Moore-Read Edge (Assuming an even number of particles in the system)}
{\begin{tabular}{c|l|l}
~~~~~$L - L_{ground}$  & ~~~~~~~~~~~~~~~states~~~\rule[15pt]{20pt}{0pt}  &    Energy  \\
\hline 
1	& $~~a^\dagger_1|0
\rangle~~~~~$ & $\phantom{1} \alpha $ \rule[10pt]{15pt}{0pt}\\
2    & $~~a^\dagger_2|0
\rangle, ~~~ a^\dagger_1 a^\dagger_1|0 
\rangle, ~~~~ \psi^\dagger_{1/2} \psi^\dagger_{3/2} | 0 \rangle$ &  
$2 \alpha $ \rule[10pt]{15pt}{0pt} \\
3    & $~~a^\dagger_3|0
\rangle, ~~~ a^\dagger_2 a^\dagger_1|0
\rangle , ~~~ a^\dagger_1 a^\dagger_1  a^\dagger_1|0
\rangle, ~~a^\dagger_1 \psi^\dagger_{1/2} \psi^\dagger_{3/2} | 0 \rangle, ~~\psi^\dagger_{1/2} \psi^\dagger_{5/2} | 0 \rangle$ &  $3 \alpha $ \rule[10pt]{15pt}{0pt} \\
 & &  \\
4 & $~~a^\dagger_4|0
\rangle, ~~~ a^\dagger_3 a^\dagger_1|0
\rangle , ~~~ a^\dagger_2 a^\dagger_1 a^\dagger_1 |0
\rangle, ~~~~ a^\dagger_2 a^\dagger_2 | 0 \rangle, ~~~~a^\dagger_1 a^\dagger_1 a^\dagger_1 a^\dagger_1 | 0 \rangle,$ &   \rule[10pt]{15pt}{0pt} \\
 &  $~~~~~a^\dagger_2 \psi^\dagger_{1/2} \psi^\dagger_{3/2} | 0 \rangle, ~~a^\dagger_1 a^\dagger_1 \psi^\dagger_{1/2} \psi^\dagger_{3/2} | 0 \rangle,~~a^\dagger_1 \psi^\dagger_{1/2} \psi^\dagger_{5/2} | 0 \rangle,$  & \\ 
&   $~~~~~~~\psi^\dagger_{1/2} \psi^\dagger_{7/2} | 0 \rangle,   
 ~~~\psi^\dagger_{3/2} \psi^\dagger_{5/2} | 0 \rangle$   
 &  $4 \alpha$ \rule[10pt]{15pt}{0pt}
\end{tabular}
}
\label{tab:mr}
\end{table}
Note that the number of edges states at a given angular momentum is greater than that of the Laughlin case shown in Table \ref{tab:cll} (for all angular momentum at least two greater than that of the ground state).  This is what we would have suspected from our above argument that we can insert polynomial factors in a greater variety of ways for the Moore-Read and Read-Rezyayi wavefunctions than for the Laughlin state. 

As in the case of the Laughlin edge, the effective Hamiltonian Eq.~\ref{eq:HMR} is exact at all energies for the special parent Hamiltonian.  Again for generic electron-electron interactions,  it is expected that at low energy the form of Eq.~\ref{eq:HMR} remains true${}^{\ref{foot:lowenergy}}$ if the ground state is in the Moore-Read phase of matter, although the two velocities for the two modes need not be the same ($\alpha$ in the two terms will not be the same). 

Analogous to the discussion of section \ref{sub:innerprod}, the different one-dimensional edge excitations (as created by products of $a^\dagger_m$ and $\psi_{m}$ operators) correspond to orthogonal two dimensional wavefunctions only in the limit of very large systems.  For finite systems, analogous to the Laughlin case, inner products of bulk states can be calculated using the exponential of a local operator $S$ which now operates on a space involving both bosonic and fermionic excitations.   See Refs.~\refcite{DubailReadRezayi} and \refcite{Fern2018A} for more details.

For the more general Read-Rezayi wavefunctions, the counting of edge excitations was achieved by Refs.~\refcite{ArdonneKedemStone} and \refcite{ReadRRCounting}, obtaining a spectrum equivalent to a so-called $\hat {su}(2)_k$ current algebra.    While this may sound fairly abstract at this point, we will see in section \ref{sec:thin}  another way to count edge excitations which does not require any knowledge of conformal field theory.

\section{Thin Limit}
\label{sec:thin}

A rather remarkable simplification occurs if one considers quantum Hall systems on a cylindrical or toroidal geometry, and takes the limit that the cylinder radius is very small.
\cite{tao1983fractional,BergholtzThinTorus1,BergholtzThinTorus2,SeidelThinTorus,thintoruslong,SeidelBilayer,SeidelLeePRB,SeidelPfaffian}
In this limit the system becomes essentially a one dimensional chain of orbitals (rings around the cylinder indexed by their angular momentum), and the clustering parent Hamiltonians $H_k^r$ discussed above, take a very simple form.   For example, the Laughlin $\nu=1/p$ special parent Hamiltonian $H_1^p$ simply gives positive energy whenever two or more particles occupy $p$ consecutive orbitals.   The resulting grounds state wavefunctions (the wavefunctions left with zero energy) in this limit are simple charge density waves\cite{tao1983fractional}.    For example the $\nu=1/3$ Laughlin state on a cylinder is given by the condition that there should be no more than one particle in any three consecutive orbitals.  Thus the ground state looks as follows:
$$
{\bf 1} \, 0 \,0 \,{\bf 1} \,0 \,0 \,{\bf 1}\, 0\, 0\, {\bf 1}\, 0\, 0\,{\bf 1}
$$
where ${\bf 1}$ indicates a filled orbital and 0 represents an empty orbital.   It is easy to see that for a finite cylinder one obtains maximum density (i.e., the ground state) only when the number of orbitals is $N_{orb} = 3 (N-1) +1$ which matches our above counting of powers of $z$, for example, in Eq.~\ref{eq:Nphiequation} with $q=1$ and $N_{\phi} + 1 = N_{orb}$.   What is not obvious (and, indeed, is only established by numerical work\cite{SeidelPfaffian,BergholtzThinTorus3,BergholtzThinTorus1}) is that this charge density wave is adiabatically connected to the Laughlin ground state.  In other words, as the thin cylinder is made thicker, the gapped charge density wave continuously deforms into the gapped Laughlin ground state. 

If we instead consider a thin torus rather than a thin cylinder, one discovers there are mulitple degenerate ground states.   For example, for Laughlin $\nu=1/3$ we have the three ground states
\begin{eqnarray*}
&
 {\bf 1}\,0\,0\,{\bf 1}\,0\,0\,{\bf 1}\,0\,0\,{\bf 1}\,0\,0 & \\
&   
 0\,{\bf 1}\,0\,0\,{\bf 1}\,0\,0\,{\bf 1}\,0\,0\,{\bf 1}\,0  & \\
&  
 0\,0\,{\bf 1}\,0\,0\,{\bf 1}\,0\,0\,{\bf 1}\,0\,0\,{\bf 1} &  \\
\mbox{\Large $\hookrightarrow$} &  & \mbox{\Large $\hookleftarrow$}
\end{eqnarray*}
where we should think of the far left as being connected up to the far right (as indicated by the hooked arrows),  and here we have $N_{orb} = 3 N$.   This ground state degeracy persists even when the torus is not thin\cite{Haldane1985periodic}.    The ground state degeneracy on the torus is characteristic of topologically ordered matter\cite{WenTopologicalOrder1,WenNiu} (indeed, it is often used as a {\it definition} of topological order).  One should be cautious, however, that topological order requires the multiple ground states to be indistinguishable from each other by any local measurement --- and this is not the case in the thin-torus limit, where one can measure the position of charges in the charge density wave.   However (and this is not an obvious statement!) as the torus is  made large in both directions the multiple ground states remain degenerate, and become locally indistinguishable.   

Analogous thin torus versions of the general clustering parent Hamiltonians $H_k^r$ follow a very similar rule:  Give positive energy to any $k+1$ particles in $r$ consecutive orbitals. 
  
As an example of this let us consider the bosonic Moore-Read state at $\nu=1$ (this is Read-Rezayi with $k=2$, having Hamiltonian $H_2^2$) with the the clustering rule that there should be no more than 2 bosons in 2 consecutive orbitals.   The ground state (the hightest density state we can make) on a cylinder is given by 
$$
   {\bf 2} \, 0 \,    {\bf 2} \, 0 \,
    {\bf 2} \, 0 \,    {\bf 2} \, 0 \,
    {\bf 2} 
$$
where ${\bf 2}$ indicates a doubly filled orbital (which is allowed since we have bosons).    Again, this gapped charge density wave ground state is continuously connected to the gapped Moore-Read ground state when the cylinder is made thicker.   It is easy to count that $N_{orb} = N_{\phi} + 1 = N -1$. 
On a torus, the clustering rule results in three different ground states given by 
\begin{eqnarray*}
&
  {\bf 2} \, 0 \,    {\bf 2} \, 0 \,
    {\bf 2} \, 0 \,    {\bf 2} \, 0 \,
    {\bf 2}  \, 0   & \\
&    
 0 \,  {\bf 2} \, 0 \,    {\bf 2} \, 0 \,
    {\bf 2} \, 0 \,    {\bf 2} \, 0 \,
    {\bf 2}    & \\
&   
 {\bf 1}\, {\bf 1}\, {\bf 1}\, {\bf 1}\, {\bf 1}\, {\bf 1}\, {\bf 1}\, {\bf 1}\, {\bf 1}\, {\bf 1} &  \\
\mbox{\Large $\hookrightarrow$} &  & \mbox{\Large $\hookleftarrow$}
\end{eqnarray*}
where again we connect up the far left to the far right, and here we have $N_{orb} = N$.    As in the Laughlin case, these states remain degenerate even when the torus is not thin.  While the three ground states are locally distinguishable in the thin  torus limit, they become indistinguishable in the large system (and thick torus) limit, signaling topological order. 

For the even more general Read-Rezayi states at $\nu=k/(2 + k p)$ the clustering rule is that there should be (a) no more than one particle in $p$ consecutive orbitals (exactly the rule for Laughlin $\nu=1/p$) and (b) no more than $k$ particles in $2+p k$ consecutive orbitals.   (Compare this to the cluster Hamiltonian prescription described just after Eq. \ref{eq:RRnu}).   As an example, we consider the fermionic $\mathbb{Z}_3$ Read-Rezayi state at filling fraction $\nu=3/5$ (this is $p=1$ and $k=3$ in Eq. ~\ref{eq:RRnu}).  Here the clustering rule is that no more than three particles are allowed in five consecutive orbitals (and no more than one particle in one orbital as required by fermionic statistics).  The ground state on the cylinder is given by  
$$
  {\bf 1} \,   {\bf 1} \,   {\bf 1} \, 0 \, 0 \, 
  {\bf 1} \,   {\bf 1} \,   {\bf 1} \, 0 \, 0 \, 
  {\bf 1} \,   {\bf 1} \,   {\bf 1} \, 0 \, 0 \, 
  {\bf 1} \,   {\bf 1} \,   {\bf 1}
$$
with $N_{orb}  = N_{\phi} + 1 = 5 N/3 -2$.  On the torus one has ten degenerate ground states given by 
\begin{eqnarray*}
&
 {\bf 1} \,   {\bf 1} \,   {\bf 1} \, 0 \, 0 \, 
  {\bf 1} \,   {\bf 1} \,   {\bf 1} \, 0 \, 0 \, 
  {\bf 1} \,   {\bf 1} \,   {\bf 1} \, 0 \, 0 \, 
  & ~~~~~~~~~~ \mbox{(and four similar translations)} \\
&    
 {\bf 1} \,   {\bf 1} \,    0 \, {\bf 1} \, 0 \, 
 {\bf 1} \,   {\bf 1} \,    0 \, {\bf 1} \, 0 \, 
 {\bf 1} \,   {\bf 1} \,    0 \, {\bf 1} \, 0 \, 
  & ~~~~~~~~~~ \mbox{(and four similar translations)} \\
\mbox{\Large $\hookrightarrow$} &  & \mbox{\Large $\hookleftarrow$}
\end{eqnarray*}
where $N_{orb} = N_{\phi} + 1 =  5 N/3$.  Unsurprisingly, the torus ground state degeneracy matches predictions from conformal field theory\cite{ArdonneTorusDomain}.

One can construct similar clustering Hamiltonians for other wavefunctions, such as the Gaffnian (the $\nu=2/3$ bosonic form having the rule of no more than two bosons in three consecutive orbitals) and the Haffnian (the bosonic $\nu=1/2$ form having the rule of no more than two bosons in four consecutive orbitals.   However, here the situation is more complicated.  While the Gaffnian and Haffnian hamiltonians may be gapped in the thin cylinder limit, they are not gapped when the cylinder is made thick\cite{PapicClustering,SeidelPert}.   Nonetheless, the states that are exactly zero energy in the thin torus limit remain exactly zero energy states as the torus is made thicker while additional states come down to zero energy only in the thermodynamic limit. 

\subsection{Edge State Counting}
\label{sub:edgestatecounting}

The thin cylinder limit for systems with simple clustering rules gives a very clean way of counting the edge modes of a quantum Hall system.   Let us begin with the Laughlin $\nu=1/3$ case where the clustering rule is that there should be no more than one filled orbital in three consecutive orbitals.   We will assume a half-infinite quantum Hall system  with the quantum Hall state existing on the far left and the vacuum existing on the far right.   The ground state is thus
$$
 \ldots  {\bf 1} \, 0 \, 0  \, {\bf 1} \, 0 \, 0  \,  {\bf 1} \, 0 \, 0 \, {\bf 1} \, 0 \, 0 \, {\bf 1} \, 0 \, 0 \, {\bf 1} \mbox{\Large $\vert$} 0 \, 0 \, 0 \, 0 \, 0 \, 0 \, 0 \, 0 \, 0 \, 0 \, \ldots   
$$
where the vertical line marks the unexcited edge of the system (the highest angular momentum occupied orbital in the ground state).   To create an edge excitation we must promote a fermion to an orbital of higher angular momentum, i.e, we move some occupied orbitals further right.   For example, if we want a state with one additional unit of angular momentum we want to promote a fermion to the the right by one more step.   There is only one way to do this without violating the clustering rule, given by
$$
\begin{array}{l}
 ~~~~~~~~~~~~~~~~~~~~~~~~~~~~~~~~~~~~~~~~~~~~~~~~~~~~~ \curvearrowright  \vspace*{-5pt}\\
 (\Delta L=1)~~~~~ \ldots  {\bf 1} \, 0 \, 0  \, {\bf 1} \, 0 \, 0  \,  {\bf 1} \, 0 \, 0 \, {\bf 1} \, 0 \, 0 \, {\bf 1} \, 0 \, 0 \, 0  \mbox{\Large $\vert$} {\bf 1} \, 0 \, 0 \, 0 \, 0 \, 0 \, 0 \, 0 \, 0 \, 0 \, \ldots   
\end{array}
$$
where we have left the vertical bar in the same place to indicate where the ground state occupation ended.   There are then two ways to make an excitation with two units of angular momentum greater than that of the ground state, given by
$$
\begin{array}{l}
  ~~~~~~~~~~~~~~~~~~~~~~~~~~~~~~~~~~~~~~~~~~~~~~~~~~~~~ \curvearrowright\! \curvearrowright  \vspace*{-5pt}\\
 (\Delta L=2)~~~~~ \ldots  {\bf 1} \, 0 \, 0  \, {\bf 1} \, 0 \, 0  \,  {\bf 1} \, 0 \, 0 \, {\bf 1} \, 0 \, 0 \, {\bf 1} \, 0 \, 0 \, 0  \mbox{\Large $\vert$}  0 \, {\bf 1} \, 0 \, 0 \, 0 \, 0 \, 0 \, 0 \, 0 \, 0 \, \ldots    \\ 
  ~~~~~~~~~~~~~~~~~~~~~~~~~~~~~~~~~~~~~~~~~~~~~~\curvearrowright  ~~ \curvearrowright   \vspace*{-5pt}\\
 (\Delta L=2)~~~~~  \ldots  {\bf 1} \, 0 \, 0  \, {\bf 1} \, 0 \, 0  \,  {\bf 1} \, 0 \, 0 \, {\bf 1} \, 0 \, 0 \,  0 \, {\bf 1} \,  0 \, 0  \mbox{\Large $\vert$} {\bf 1} \, 0 \, 0 \, 0 \, 0 \, 0 \, 0 \, 0 \, 0 \, 0 \, \ldots    
\end{array}
$$
In other words, we may promote the furthest right fermion by two orbitals, or we may promote the two furthest right fermions each by one orbital.    For excitations three units of angular momentum greater than that of the ground state, there are three possibilities (promote the furthest right fermion by three orbitals;  promote the furthest right fermion by two orbitals and the next furthest right fermion by one orbital; promote the three furthest right fermions each by one orbital).  Comparing this counting to that of Tables \ref{tab:cll} and  \ref{tab:sympoly} we can see that we are describing exactly the same counting (i.e., partitions of integers).

We can also apply the thin cylinder limit to counting edge excitations for more complicated wavefunctions, such as Moore-Read, Read-Rezayi\cite{BergholtzReview,BergholtzThinTorus1,
BergholtzThinTorus2,SeidelLeePRB,SeidelPfaffian,
SeidelThinTorus,ArdonneDomain}, and even Gaffnian, Haffnian\cite{SeidelPert,PapicClustering} or bilayer wavefunctions\cite{SeidelBilayer}.   As an example, let us consider the case of the Moore-Read state of bosons at $\nu=1$.  Here, the clustering rule is that there should be no more than 2 bosons in two consecutive orbitals.  For a half-infinite cylinder we consider the ground state to be 
$$
~~~~~~~~~~~~~~~~~  \ldots {\bf 2} \, 0 \,    {\bf 2} \, 0 \,     {\bf 2} \, 0 \,    {\bf 2} \, 0 \,
    {\bf 2} \, 0 \,    {\bf 2} \, 0 \,
    {\bf 2} \mbox{\Large $\vert$} 0 \, 0 \, 0 \, 0 \, 0 \, 0 \, 0 \, 0 \, 0 \, \ldots 
$$
where again, we use the vertical line to indicate the position of the furthest right (highest angular momentum) occupied orbital in the ground state.   Let us list off the possible edge excitations in order of their angular momentum. There is a single state with angular momentum one unit greater than the ground state
\begin{eqnarray*}
(\Delta L = 1) ~~~~~ & & \ldots {\bf 2} \, 0 \,    {\bf 2} \, 0 \,     {\bf 2} \, 0 \,    {\bf 2} \, 0 \,
    {\bf 2} \, 0 \,    {\bf 2} \, 0 \,
    {\bf 1} \mbox{\Large $\vert$} {\bf 1} \, 0 \, 0 \, 0 \, 0 \, 0 \, 0 \, 0 \, 0 \, \ldots  
\end{eqnarray*}
There are three wavefunctions with angular momentum two units greater than the ground state
\begin{eqnarray*}        
    (\Delta L = 2) ~~~~~ & & \ldots {\bf 2} \, 0 \,    {\bf 2} \, 0 \,     {\bf 2} \, 0 \,    {\bf 2} \, 0 \,
    {\bf 2} \, 0 \,    {\bf 2} \, 0 \,
     {\bf 1} \mbox{\Large $\vert$} 0 \, {\bf 1} \,  0 \, 0 \, 0 \, 0 \, 0 \, 0 \, 0 \, \ldots      \\ 
(\Delta L = 2) ~~~~~ & & \ldots {\bf 2} \, 0 \,    {\bf 2} \, 0 \,     {\bf 2} \, 0 \,    {\bf 2} \, 0 \,
    {\bf 2} \, 0 \,    {\bf 2} \, 0 \,
     0 \mbox{\Large $\vert$} {\bf 2} \, 0 \, 0 \, 0 \, 0 \, 0 \, 0 \, 0 \, 0 \, \ldots     \\
    (\Delta L = 2) ~~~~~ & & \ldots {\bf 2} \, 0 \,    {\bf 2} \, 0 \,     {\bf 2} \, 0 \,    {\bf 2} \, 0 \,
    {\bf 2} \, 0 \,    {\bf 1} \, {\bf 1} \,
     {\bf 1} \mbox{\Large $\vert$}  {\bf 1} \, 0 \,  0 \, 0 \, 0 \, 0 \, 0 \, 0 \, 0 \, \ldots       
\end{eqnarray*}
and there are five wavefunctions with angular momentum three units greater than the ground state
\begin{eqnarray*}     
   (\Delta L = 3) ~~~~~ & & \ldots {\bf 2} \, 0 \,    {\bf 2} \, 0 \,     {\bf 2} \, 0 \,    {\bf 2} \, 0 \,
    {\bf 2} \, 0 \,    {\bf 2} \, 0 \,
     {\bf 1} \mbox{\Large $\vert$} 0 \,   0 \, {\bf 1} \, 0 \, 0 \, 0 \, 0 \, 0 \, 0 \, \ldots      \\   
        (\Delta L = 3) ~~~~~ & & \ldots {\bf 2} \, 0 \,    {\bf 2} \, 0 \,     {\bf 2} \, 0 \,    {\bf 2} \, 0 \,
    {\bf 2} \, 0 \,    {\bf 2} \, 0 \,
     0 \mbox{\Large $\vert$} {\bf 1} \,   {\bf 1} \, 0   \, 0 \, 0 \, 0 \, 0 \, 0 \, 0 \, \ldots      \\       
     {}
        (\Delta L = 3) ~~~~~ & & \ldots {\bf 2} \, 0 \,    {\bf 2} \, 0 \,     {\bf 2} \, 0 \,    {\bf 2} \, 0 \,
    {\bf 2} \, 0 \,    {\bf 1} \, {\bf 1}  \,
      {\bf 1} \mbox{\Large $\vert$} 0 \,   {\bf 1} \, 0   \, 0 \, 0 \, 0 \, 0 \, 0 \, 0 \, \ldots      \\              
      {}
(\Delta L = 3) ~~~~~ & & \ldots {\bf 2} \, {\bf 0} \,    {\bf 2} \, 0 \,     {\bf 2} \, 0 \,    {\bf 2} \, 0 \,
    {\bf 2} \, 0 \,    {\bf 1} \, {
    \bf 1} \,
     0 \mbox{\Large $\vert$} {\bf 2} \, 0 \, 0 \, 0 \, 0 \, 0 \, 0 \, 0 \, 0 \, \ldots     \\
{}  
(\Delta L = 3) ~~~~~ & & \ldots {\bf 2} \, {\bf 0} \,    {\bf 2} \, 0 \,     {\bf 2} \, 0 \,    {\bf 2} \, 0 \,
    {\bf 1} \, {\bf 1} \,    {\bf 1} \, {\bf 1} \,
     {\bf 1} \mbox{\Large $\vert$} {\bf 1} \, 0 \, 0 \, 0 \, 0 \, 0 \, 0 \, 0 \, 0 \, \ldots     
  \end{eqnarray*}
We see that this counting matches up precisely with our description of a bosonic and a majorana fermionic edge mode as given in Table \ref{tab:mr}.   With a bit of work one can show that the correspondence continues at higher angular momentum as well\cite{SeidelPfaffian,ReadRRCounting,ArdonneDomain,BergholtzThinTorus2} .

This type of thin cylinder state counting is extremely simple.  For more general clustering Hamiltonians, including those for the Read-Rezayi series, the Haffnian and the Gaffnian, it can be shown that this type of thin cylinder edge state counting precisely matches the predictions of the corresponding conformal field theory\cite{ArdonneDomain}.

\section{Squeezing and Jack Polynomials}
\label{sec:Jack}

The thin cylinder limit is advantageous because of the simplicity of the resulting wavefunctions.   In the thin limit, each eigenstate is a single Fock state (a single list of occupations) rather than a superposition of many Fock states.    When we make the cylinder thick again the eigenstates are not single Fock states but are rather superpositions.  Nonetheless some of the structure of the thin cylinder remains. 

For simplicity in this section let us assume we are thinking about a system of bosons and we are considering clustering Hamiltonians $H_k^r$ which enforces the constraint that the wavefunction must vanish as at least $r$ powers when $k+1$ particles come to the same point.   For the case of $r=2$ these produce the $\mathbb{Z}_k$ Read-Rezayi wavefunctions of the form of Eq.~\ref{eq:RRwavefunction} (without additional Jastrow factors) as a ground state.  

In the limit of a thin cylinder $H_k^r$ imposes a clustering rule that there should be no more than $k$ bosons in $r$ consecutive orbitals.    Let us define the Fock states that satisfy  this $(k,r)$ clustering condition of the thin cylinder to be called $(k,r)$-admissable.    For thick cylinders (or on the plane or sphere) these $(k,r)$-admissable Fock states are not eigenstates.  However, each admissible Fock state is nonetheless in one-to-one correspondence with a zero energy eigenstates of the corresponding clustering Hamiltonian $H_k^r$. 

To generate a basis of zero energy eigenstates we can use each $(k,r)$-admissable state as a so-called {\it root}.  Then each root can be superposed with so-called  descendant Fock states\cite{HaldaneBAPS} to create eigenstates.   A descendant state is defined from a root state by an operation known as {\it squeezing} where two particles are moved towards each other by one step in angular momentum.  We sometimes say that if one Fock state descends from another we say it is {\it dominated} by the other state.  The descendant states may or may not be admissable.   As an example, let us consider one of the root states describing one of the $\Delta L=3$ edge excitations of the Moore-Read state, and list off some of its descendants.  The root state must be $(k,r)$ admissable, which in this case is $(2,2)$ admissable:  
$$
\begin{array}{ll}
     (\mbox{root-admissable}) ~~~~~ &  \ldots {\bf 2} \, {\bf 0} \,    {\bf 2} \, 0 \,     {\bf 2} \, 0 \,    {\bf 2} \, 0 \,
    {\bf 2} \, 0 \,    {\bf 1} \, {
    \bf 1} \,
     0 \mbox{\Large $\vert$} {\bf 2} \, 0 \, 0 \, 0 \, 0 \, 0 \, 0 \, 0 \, 0 \, \ldots           \\
     \downarrow \mbox{squeeze} & ~~~~~~~~~~~~~~~~~~~~~\curvearrowright  ~~~~ \curvearrowleft \vspace*{-2pt}\\
    (\mbox{admissable}) ~~~~~ &  \ldots {\bf 2} \, {\bf 0} \,    {\bf 2} \, 0 \,     {\bf 2} \, 0 \,    {\bf 2} \, 0 \,
    {\bf 1} \, {\bf 1} \,    {\bf 1} \, {
    \bf 1} \,
     {\bf 1} \mbox{\Large $\vert$} {\bf 1} \, 0 \, 0 \, 0 \, 0 \, 0 \, 0 \, 0 \, 0 \, \ldots           \\
     \downarrow \mbox{squeeze} & ~~~~~~~~~~~\, ~\curvearrowright  ~~~~~~~~~~ \curvearrowleft \vspace*{-2pt}\\
    (\mbox{non-admissable})   &  \ldots {\bf 2} \, {\bf 0} \,    {\bf 2} \, 0 \,     {\bf 1} \, {\bf 1} \,    {\bf 2} \, 0 \,
    {\bf 1} \, {\bf 1} \,    {\bf 1} \, {
    \bf 2} \,
      0 \mbox{\Large $\vert$} {\bf 1} \, 0 \, 0 \, 0 \, 0 \, 0 \, 0 \, 0 \, 0 \, \ldots    \\ 
     \downarrow \mbox{squeeze} & ~~~~~~~~~~~~~~\curvearrowright  ~~~~~~~~~~~ \curvearrowleft \vspace*{-2pt}\\
    (\mbox{non-admissable})   &  \ldots {\bf 2} \, {\bf 0} \,    {\bf 2} \, 0 \,     {\bf 1} \, 0 \,    {\bf 3} \, 0 \,
    {\bf 1} \, {\bf 1} \,    {\bf 1} \, {
    \bf 2} \,
     {\bf 1} \mbox{\Large $\vert$} 0 \, 0 \, 0 \, 0 \, 0 \, 0 \, 0 \, 0 \, 0 \, \ldots    \\      
 \end{array}
 $$
To form an eigenstate for a thick cylinder (or sphere or disk) one needs to superpose a root state with its descendants with the proper coefficients.     Just knowing that an eigenstate is comprised only of a root and its descendants greatly reduces the size of the Hilbert state that one is considering and this is extremely powerful for numerical work\cite{HaldaneBAPS}. 

The determination of the exact coefficients of the Fock states, and hence the wavefunction itself,  appears to be a complicated problem.   We must somehow figure out how to superpose the root with all the descendant states in such a way so as to satisfy the clustering Hamiltonian on the non-thin-cylinder geometry.  (For the bosonic Read-Rezayi states this would be finding the wavefunction that vanishes as $r$ powers when $k+1$  particles come to the same point.)  Rather incredibly the answer to this problem is given by {\it Jack Polynomials}\cite{BernevigHaldane1}
$$
 \Phi^{\nu=k/r; \mbox{\scriptsize bosons}} =  J^\alpha_{\mbox{\boldmath$\lambda$}}(z_1, \ldots, z_N)
$$
where $\alpha = -(k+1)/(r-1)$ and {\boldmath$\lambda$} is a $(k,r)$-admissable root state (in usual Jack notation this root state is presented as a partition of an integer\cite{BernevigHaldane1} although in the FQHE world it is more convenient to specify this partition as an admissable Fock state).    An enormous amount is known about these polynomials\cite{Jack1970,Stanley1989} and a recent work from the mathematical literature\cite{Feigin} found that for this value of $\alpha$, at least when $k$ and $r$ are coprime, the resulting polynomials have precisely the desired vanishing properties --- that the wavefunction vanishes as $r$ or more powers when $k+1$ particles approach the same position.   These Jacks provide explicit wavefunctions for the ground state, corresponding to filling fraction $\nu=k/r$ (as in Eq.~\ref{eq:nukr}), as well as providing explicit wavefunctions that span the space of all the edge excitations (i.e., all lower density wavefunctions that also have zero interaction energy).   The Jacks describe  the $\mathbb{Z}_k$ Read-Rezayi states $(k,r) =(k,2)$ including the Laughlin ($k=1$) and Moore-Read ($k=2$) states, as well as the Gaffnian wavefunction $(k,r) = (2,3)$ and a host of other wavefunctions which have now been identified\cite{Bernevig2009C,Estienne2009} as being described by so-called $WA_{k-1}(k+1,k+r)$ conformal field theories.   

For both the bosonic version and the fermionic version (with one Jastrow factor attached) of these wavefunctions, the mathematical structure of the Jack polynomials obey interesting rules whereby wavefunctions on large systems can be constructed efficiently by ``sewing" together  wavefunctions from small systems\cite{Bernevig2009B,Estienne2011}.  These algorithms have allowed numerical work on systems that are far larger than would otherwise be possible.  Indeed, the use of Jacks has become an indespensible part of the numerical FQHE toolbox. 

The rich mathematical structure of the Jacks has allowed detailed analysis of these (and other) wavefunctions in a host of new ways, both numerical and analytical\cite{BernevigHaldane1,BernevigHaldane2,
BernevigHaldane3,BernevigHaldane4,Bernevig2009C,
Estienne2009,Estienne2010A,Estienne2010C,
Estienne2011,
WanJackEdge,WojsJack}.   
The Jack approach has been extended, not only to fermionic systems, but also to spin-singlet wavefunction\cite{Bernevig2009B,Estienne2011,Estienne2012}.

As an interesting aside, we point out that Jack polynomials  $J^\alpha_{\mbox{\boldmath$\lambda$}}$ with different indices $\alpha=1/p$ turn out to describe the eigenstates of Laughlin $\nu=1/p$ droplets in an extremely steep confining potential\cite{Fern2017}. 

\section{Entanglement}
\label{sec:entanglement}

In the last decade condensed matter theory has increasingly turned to
methods of quantum information for the understanding of interacting
systems (See for example Ref.~\refcite{ChenWenBook}).  A particularly
valuable approach is to partition a system into two pieces (a
``bipartition") and to examine the quantum entanglement between the
pieces\cite{Eisert}.  This approach has proven to be extremely
valuable in the study of quantum Hall effect as well.  

One can generally partition the Hilbert space ${\cal H}$ of a system into two pieces
$$
 {\cal H} = {\cal H}^A \otimes {\cal H}^B
$$
This division can be made in a number of different ways.  For example, one might put certain orbitals of a system in $A$ and  the other orbitals in $B$ (known as an ``orbital partition"\cite{LiHaldane}).  Another possibility is that certain particles are in $A$ and the others are in $B$ (a ``particle partition"\cite{Zozulya}).  Yet another possibility is to partition the area of a system into two sub-areas ($a$ and $b$) and put particles in $A$ if they fall in area $a$ and into $B$ if they fall in area $b$ (a ``real-space partition"\cite{Dubail1,RodriguezRealSpace,SterdyniakRealSpace}). 

A wavefunction $|\Psi\rangle$ for the full system may be written in terms of its partitions using a Schmidt decomposition
\begin{equation}
 |\Psi \rangle  = \sum_n \, \lambda_n \,\, |\psi^A_n\rangle \otimes |\psi^B_n\rangle
 \label{eq:Schmidt}
\end{equation}
where the wavefunctions $|\psi^A_n\rangle$ form an orthonormal complete set for the Hilbert space $A$ and the wavefunctions $|\psi^B_n\rangle$ form an orthonormal complete set for the Hilbert space $B$.

Initial studies\cite{Zozulya,haque2007entanglement,IvanSierra,FradkinNowling} of the entanglement of quantum Hall systems focused on the von Neumann  entropy of entanglement given by 
$$
 S_{AB} = - \sum_n \, \lambda_n \log \lambda_n
$$
In cases where $A$ and $B$ spatially partition a system into two pieces, the entanglement entropy should have the behavior 
$$
  S_{AB} = \alpha L + D  + \ldots
$$
where $L$ is the length of the cut between the two pieces, $\alpha$ is a nonuniversal constant, and $D$ is a subleading piece (known as the total quantum dimension) which encodes information about the topological properties of the phase of matter\cite{LevinWenEntanglement,KitaevPreskill}.

In Ref.~\refcite{LiHaldane} it was emphasized that the entire spectrum of Schmidt weights $\lambda_n$  in Eq.~\ref{eq:Schmidt} contains important information and most of this is thrown away by looking only at the von Neumann entanglement entropy.  The full set of Schmidt weights is known as the ``entanglement spectrum". 

The information contained in the entanglement spectrum is perhaps most clear when we partition the system into pieces $A$ and $B$ so as to conserve certain additional quantum numbers besides particle number $N$, such as angular momentum $L$.   In this case, the Hilbert space has a  ``graded" structure
$$
 {\cal H}_{N,L} = \bigoplus_{N_A}  \bigoplus_{L_A} ~~~~~ {\cal H}^A_{L_A,N_A} \otimes {\cal H}^B_{L_B,N_B} 
$$
where $N=N_A + N_B$ and $L=L_A + L_B$.    Each term  in the Schmidt decomposition must then be similarly labeled with these quantun numbers.    For example, we might have a Schmidth weight $\lambda_n(N^A,L^A)$ meaning that in the product $|\psi^A_n\rangle \otimes |\psi^B_n\rangle$, piece $A$ has $N^A$ particles and angular momentum $L^A$.   

Most frequently one considers bipartitioning a sphere in a manner which is symmetric under rotations around the north-south axis (i.e., making a cut along a lattitude) so that the $L_z$ angular momentum is conserved by the cut.   A similar possibility is cutting a cylinder around an equator.    Let us for now consider either an orbital or real-space partition of this sort which conserves angular momentum.     Fixing $N_A$, the number of particles that end up on the $A$ side of the cut, the  entanglement spectrum is conventionally plotted by showing 
$$ 
\xi = -2 \log \lambda
$$
as a function of $L_A$ the angular momentum of the particles on the $A$ side of the cut.  The values $\xi(L)$ are often referred to as the ``entanglement energies" and the reduced density matrix 
$$
 \rho_A = \sum_n \lambda_n |\phi^A_n\rangle \langle \phi^A_n|  
$$
can be written as $\rho_A = e^{-H_E}$ with $H_E$ being the so-called ``entanglement Hamiltonian" whose eigenvectors are $|\phi^A_n\rangle$ and corresponding eigenvalues are $\xi_n$. 

Rather remarkably,  the entanglement spectrum that results from this procedure has the same structure as the energy spectrum of the edge of the given quantum Hall state\cite{LiHaldane,RegnaultReview}.  In other words, the entanglement Hamiltonian $H_A$ is similar to the Hamiltonian of a system having $N_A$ particles in a confining potential --- having linearly dispersing edge modes of low energy excitations in the long wavelength limit.     

For example, for wavefunctions having special parent Hamiltonians, such as the Laughlin wavefunction (or the Read-Rezayi states or any of the other $\nu=k/r$ Jack wavefunctions), the counting of states in the edge spectrum (the number of modes at any given angular momentum $L$) precisely matches that of the  entanglement spectrum.     Such countings (so-called ``entanglement-spectroscopy") are now commonly used as a way to identify a particular phase of matter from exact diagonalization\cite{RegnaultReview}.    

One can also consider wavefunctions from more realistic Hamiltonians.  For example, considering a Coulomb interaction between electrons one finds the ground state at $\nu=1/3$ to be a quantum Hall state, presumably in the Laughlin phase of matter.  The corresponding ground state wavefunction of the Coulomb interaction may be extremely similar to the Laughlin wavefunction, but will have some small differences.   In the entanglement spectrum, there will be some modes with low entanglement energy which match that of the Laughlin wavefunction, but at high entanglement energy (smaller $\lambda$ weight in the Schmidt decomposition) there are additional modes indicating some deviation from the ideal wavefunction\cite{Sterd2}.   A so-called ``entanglement gap" separates the low entanglement energy expected modes and the high entanglement energy additional modes.  

One can also consider the entanglement spectra of wavefunctions that do not have a simple parent Hamiltonian or simple Jack description.  For example, one can consider the entanglement spectrum of a Jain wavefunction\cite{RodriguezEntanglementCF,RegnaultClustering,DavenportEntanglementCF}.  As one might expect, one obtains a spectrum corresponding to multiple bosonic modes which matches the presumed edge spectrum, along with extra modes with small weight above an entanglement gap. 

There are several discussions of why the edge spectrum and the entanglement spectrum have such similar structure\cite{Dubail1,DubailReadRezayi,QiKatsura,Swingle}.  Although none of these arguments is particularly simple, the approach by Ref.~\refcite{QiKatsura} is useful to discuss as it introduces some general ideas (although many of the same ideas reoccur in different form in Ref.~\refcite{DubailReadRezayi}).   One considers splitting the Hamiltonian into the part acting on each part of the partition $H_A$ and $H_B$ as well as a coupling between the two pieces $H_{AB}$ 
$$
 H = H_A + H_B + \lambda H_{AB.}
$$

If we consider a case where $\lambda$ is small (and assuming the system is everywhere gapped) then one can qualitatively understand the structure.  If we consider $H_A$ only, acting on a fixed number of particles $N_A$ which are in the $A$ part of the system, we basically have a quantum Hall system with a gapless edge (a quantum Hall droplet).  Similarly on the other side of the boundary with $H_B$ we have a system with a gapless edge.    The small Hamiltonian $\lambda H_{AB}$ cannot effect the states in the bulk of either the $A$ or $B$ side, since the bulk is gapped.   However, $\lambda H_{AB}$ couples (and generically will entangle) the otherwise gapless edge modes of the two sub-systems.  This immediately gives us the correct number of states in the entanglement spectrum.  I.e., the counting of states at each angular momentum matches between the entanglement spectrum and the edge state spectrum. 
  
In some more detail, the two chiral systems with boundary can be thought of as a single achiral system with boundary --- and with renormalization group arguments, one  can employ ideas of boundary conformal field theory.  In particular, at long length scale it is argued that the system must have conformal invariance.  It is well known that for rational conformal field theories, the only possible conformally invariant boundary states are so-called Ishibashi states\cite{Ishibashi} which entangle only states with complementary quantum numbers, in fact giving a maximally entangled state\cite{QiKatsura,DubailReadRezayi}.   These Ishibashi states have a completely flat entanglement spectrum.  This is the fixed point of the renormalization procedure.  For a system of finite size, one needs to move back away from the fixed point and examine how the Schmidt weights vary.    The leading term is given by a simple linear spectrum in angular  momentum
\begin{equation}
 H_{E} = \mbox{constant} +\kappa L +  \ldots
\end{equation}
entirely analogous to a linear spectrum of a confined edge where the constant $\kappa$ is inversely proportional to the length of the boundary and depends also on the magnitude of the coupling $\lambda$ between the two sides.    Subleading terms are smaller in orders of the length of the boundary (i.e, in powers of $1/\sqrt{N}$). 

In Ref.~\refcite{DubailReadRezayi} the analysis is taken somewhat further.  Focusing on the case where the Hamiltonian is uniform across the entire system, and on cases where there is a special parent Hamiltonian,   one can use the ideas of writing the edge state inner products as the exponential of a local operator, as in section \ref{sub:innerprod}, to determine the entanglement Hamiltonian --- which analogously can be expanded in terms involving only local operators of an appropriate field.   Indeed, the expansion of the subleading terms in the entanglement Hamiltonian is identical to the expansion of the local operator that determines the inner product between bulk wavefunctions as discussed in section \ref{sub:innerprod} above.

\section{Localized Quasiholes}
\label{sec:quasiholes}

After this extensive discussion of edge excitations of model quantum Hall states, we turn to the question of localized excitations in the bulk.  There are naturally two types of low energy ``quasiparticle" excitations: The ``quasielectron", a minimal localized packet of charge with the same sign as that of the underlying electrons, and the ``quasihole", a minimal localized packet of charge with the opposite sign.    If we continue to think about the case of simple special parent Hamiltonians, the quasiholes are particularly simple as they have zero interaction energy, like the edge states we have already considered.  Indeed, given that our definition of the edge excitations spans the entire space of possible wavefunctions with zero interaction energy, the quasiholes live within the space of what we termed edge excitations.    Unless they happen to sit at a particularly symmetric point, like the center of a disk, or a pole of the sphere, the wavefunctions in the presence of quasiholes are not generally angular momentum eigenstates, but are rather superpositions of many different angular momenta.   To see how this works, let us consider the case of the well-known Laughlin quasihole
\begin{equation}
 \Phi_{qh}(w; \,  z_1, \ldots, z_N) = \left[ \prod_{i=1}^N (z_i -w) \right] \Phi_{\mbox{\scriptsize Laughlin}}(z_1, \ldots z_N) 
 \label{eq:laughlinqh}
\end{equation}
The physics here is that the quasihole prefactor 
pushes charge away from the point $w$.   Indeed, 
pushing this minimal unit of charge away from the point $w$ is equivalent to Laughlin's adiabatic flux insertion argument\cite{laughlintheory}.   Since insertion of a flux quantum pushes $\nu$ units of electron charge away from the flux, we can conclude that the quasihole charge is $-\nu$ times the charge of the underlying electron. 

Note that this form of wavefunction is also a polynomial times the Laughlin ground state and is therefore a state with zero interaction energy.    However, this polynomial is not homogeneous in degree, and therefore is not an angular momentum eigenstate.   Instead, the quasihole factor includes all possible angular momenta in a superposition chosen in such a way that a hole is localized at a single point.  To see how this happens,  let us expand the quasihole factor into its angular momentum components
$$
\prod_i (z_i -w) = (-w)^N   + e_1 (-w)^{N-1} + e_2 (-w)^{N-2} + \ldots e_{N-1} )(-w) + e_N
$$
where $e_j$ is the symmetric monomoial
$$
 e_j =  \sum_{i_1 < i_2 < \ldots i_j} z_{i_1} z_{i_2} \ldots z_{i_j}
$$
Thus, the quasihole is a ``coherent" superposition of  many edge state excitations.  In this expansion no $z_i$ ever occurs with greater than one power --- thus the wavefunction corresponds to increasing $N_\phi$ by exactly one flux quantum.   Note that the component of the wavefunction including $e_j$ has $j$ units of angular momentum (i.e.,  $j$ powers of $z$'s) added to the Laughlin ground state.   Further, the factors of $w^p$ represent the ``wavefunction" of the quasihole.  Analogously to having an electron in a $z^p$ orbital, this puts the quasihole in the $w^p$ orbital.    Similar coherent state constructions of localized quasiholes have also been achieved for other Jack ground state wavefunctions (including Moore-Read, Read-Rezayi, Gaffnian, etc) in terms of superpositions of Jack polynomials\cite{BernevigHaldane3}.

A more direct way to understand the quasiholes of quantum Hall states is via explicit wavefunction expressions.   As an example, let us consider the Moore-Read wavefunction written in the form of Eq.~\ref{eq:PfEq}.   Here we can insert two quasiholes, one at position $w$ and another at position $w'$, by rewriting the matrix factor of Eq.~\ref{eq:PfM} representing the pairing wavefunction as\cite{Moore89}
\begin{equation}
\label{eq:Mijz}
 M_{ij} = \frac{(z_i - w)(z_j - w')}{z_i - z_j}
\end{equation}
and we write a wavefunction for all the electrons in the presence of the quasiholes\footnote{Usually a Pfaffian is defined only for antisymmetric matrices, so we should antisymmetrize $M_{ij}$ before taking the Pfaffian.} as $\Phi={\rm Pf}({\bf M})\prod (z_i - z_j)$ as in Eq.~\ref{eq:PfEq}.   In Eq.~\ref{eq:Mijz} one particle of the $z_i,z_j$ pair sees a zero of the wavefunction at position $w$ and the other sees a zero of the wavefunction at position $w'$.  Of course in the end the wavefunction is fully (anti)symmetric for (fermions)bosons.  It is easy to check that this wavefunction remains a zero energy eigenstate of the three-body special parent Hamiltonian (ex., for bosons the wavefunction vanishes when three particles come to the same point).   Note that if the two quasihole coordinates coincide, $w=w'$ this would be equivalent to multiplying the original Moore-Read wavefunction by a Laughlin quasihole factor similar to Eq.~\ref{eq:laughlinqh}, and thus would amount to a compound quasihole of total charge $-\nu$ times the charge of the underlying electron.      This charge is split into two when $w$ and $w'$ are moved apart from each other. Thus, each such quasihole has charge $-\nu/2$ times the charge of the constituent ``electron".   This should fit with our picture where only half of the electrons see a zero of the wavefunction at each quasihole coordinate. 

One can add more quasiholes similarly by writing\cite{nayakwilczek}
\begin{equation}
 M_{ij} = \frac{\prod_{a=1}^M (z_i - w_a)(z_j - w_a')}{z_i - z_j}
\label{eq:NWform}
\end{equation}
which then has $2M$ quasiholes.   For some fixed set of $2M$ quasihole coordinates, one then has a choice of which of these coordinates to put in the $w$ group and which to put in the $w'$ group.  
This gives a large number of different possible quasihole wavefunctions.  In fact we have $\frac{1}{2}(2M$ choose $M)$ = $\frac{1}{2}(2M)!/(M!)^2$ different wavefunctions corresponding to the different possible choices.   Crucially, only $2^{M-1}$ of these wavefunctions are linearly independent in agreement with predictions from conformal field theory\cite{nayakreview,ReadRezayi1996,moore1991nonabelions,ReadGreen}. A very nontrivial statement is that for a large system with many electrons, and where all of the quasiholes are far apart from each other, there is no local measurement that can distinguish these different wavefunctions from each other.   This statement, while not proven, is believed to be true and is supported by both numerical\cite{BarabanThesis,tserkovnyak,baraban,wuestienne} and theoretical\cite{ReadGreen,bondersongurarie} work.  It is this fact that is the essence of nonabelian statistics, which we will discuss further in section \ref{sub:nonabelian} below.

A different form of the Moore-Read wavefunction with quasiholes is given by the generalization of Eq.~\ref{eq:RRwavefunction}.  (Recall that the $k=2$ case of the Read-Rezayi wavefunction is the Moore-Read wavefunction.)   Here we add quasiholes by adding factors similar to that of the Laughlin quasiholes (as in Eq.~\ref{eq:laughlinqh}) to each group of particles in Eq.~\ref{eq:RRwavefunction} to obtain\cite{cappelli2001parafermion}
\begin{eqnarray} \nonumber
 \Phi_{\mbox{\scriptsize Moore-Read with qh's}}^{\nu=1/2; \mbox{ \scriptsize bosons}} &=& {\mathbb S}\left[  \left(\prod_{a=1}^M  \,\, \prod_{0 < s \leq N/2} (z_s - w_{a}) \prod_{0<i<j \leq N/2}(z_i - z_j)^2 \right) \right.
 \\  
   && \left.  ~~ \left( \prod_{a=1}^M  \,\, \prod_{N/2 < s \leq N} (z_s - w_{a}') \prod_{N/2<i<j \leq N}(z_i - z_j)^2 \right)  \right]   
\end{eqnarray}
Again it is easy to check that this wavefunction is a zero energy state of the special parent Hamiltonian (the wavefunction must vanish when three particles come to the same point, since two of the particles must be in the same group).   Similar to the case of Eq.~\ref{eq:NWform} we have a choice of which quasihole coordinates to assign to the $w$ group and which to the $w'$ group.  In fact again we have $\frac{1}{2} (2M$ choose $M)$ different possible choices, which span the $2^{M-1}$ dimensional space.    

This form of wavefunction can easily be generalized to the entire Read-Rezayi series\cite{cappelli2001parafermion}.    For the $\mathbb{Z}_k$ Read-Rezayi state with $k M$ quasiholes, we divide the quasihole coordinates into $k$ groups and write
\begin{eqnarray} \nonumber
& &  \Phi_{\mbox{\scriptsize Read-Rezayi with qh's}}^{\nu=k/2; \mbox{ \scriptsize bosons}} = {\mathbb S}\left[  \left(\prod_{a=1}^M  \,\, \prod_{0 < s \leq N/k} (z_s - w_{a,1}) \prod_{0<i<j \leq N/k}(z_i - z_j)^2 \right) \right.
 \\  & &  ~~~~~~~~~~~~~\nonumber
 \left( \prod_{a=1}^M  \,\, \prod_{N/k < s \leq 2N/k} (z_s - w_{a,2})  \prod_{N/k<i<j \leq 2N/k}(z_i - z_j)^2 \right)  \ldots  \\
   &\ldots & \left.   \left( \prod_{a=1}^M  \,\, \prod_{N (k-1)/k < s \leq N} (z_s - w_{a,k}) \prod_{N(k-1)/k<i<j \leq N}(z_i - z_j)^2 \right)  \right]   
\end{eqnarray}
so that in the $p^{th}$ group of particles we have inserted $M$ quasiholes at positions $\{ w_{a,p} \}$ for $a=1 \ldots M$.   Each $w_{a,p}$ position corresponds to a local depression in density.  However, only the $z$ coordinates in the $p^{th}$ group see a zero of the wavefunction at position $w_{a,p}$ for any $a$.  As with the Moore-Read states there  is a huge freedom to choose which coordinates to put in which group.  And again many of the resulting wavefunctions are not  linearly independent.    Using a slightly different form of the wavefunction the number of linearly independent wavefunctions is enumerated in Ref.~\refcite{ReadRRCounting}.   As in the Moore-Read case, when the quasiholes are sufficiently far apart from each other the different linearly independent wavefunctions are presumed to be indistinguishable by any local measurement.

\subsection{Nonabelian Statistics in Brief}
\label{sub:nonabelian}

The ideas of non-abelian statistics have been reviewed elsewhere\cite{nayakreview,kitaev2003,Kitaev20062}.  However, for completeness it is worth describing the basic idea here. 

Let us consider any of the $\mathbb{Z}_k$ Read-Rezayi states for $k>1$, and let us suppose we now have some number $M$ of quasiparticles\footnote{It is simplest to assume quasiholes with a special parent Hamiltonian so that the energy is simply zero.  However, most of the argument is unchanged for quasielectrons and even without a fine-tuned Hamiltonian.} at well separated positions $w_1 \ldots w_M$.  There is a multi-dimensional space of degenerate states which can describe quasiparticles at these positions (let us call this dimension $D$).   As we claimed above, the states of this space are indistinguishable by any local measurements. 
 Given any orthonormal basis for this space 
$|\Psi_n(w_1, \ldots, w_M)\rangle$, a generic wavefunction is some linear combination
$$
 |\Psi(w_1, \ldots, w_M) \rangle = \sum_{n=1}^D A_n  |\Psi_n(w_1, \ldots, w_M) \rangle
$$
with arbitrary complex coefficients $A_n$ subject to normalization $\sum_n |A_n|^2 =1$.   

Now we consider moving the particles adiabatically.   We can imagine that the quasiparticles are attracted to microscopic trapping potentials (small charges) and these traps are slowly moved.   Suppose now that at the end of this motion the quasiparticles are  again at the same positions  $w_1, \ldots w_M$.  The wavefunction must again be assembled from the same orthonormal basis, so we write
$$
 |\Psi'(w_1, \ldots, w_M) \rangle = \sum_{n=1}^D A_n'  |\Psi_n(w_1, \ldots, w_M) \rangle
$$
and we must have
\begin{equation}
\label{eq:Umatrix}
  A_m'  =  \sum_{n} U_{m n} A_n 
\end{equation}
where $U_{mn}$ is a unitary $D$ by $D$  matrix which  depends only on the path the particles have taken (assuming the eigenstate energy is always zero\footnote{This assumption is appropriate for quasiholes with a special parent Hamiltonian.  Without this assumption there is an additional abelian phase $\exp (i \int E(t) dt)$.}).  More to the point, up to an abelian phase 
the unitary matrix depends only on the topology of the path  --- i.e., on what knot has been formed by the space-time world lines.     

The controlled application of unitary matrices to a given Hilbert space is most of what is required in order to build a quantum computer.  All we really need in addition is the ability to initialize and measure the system.  The idea of using such braiding to build a quantum computer is known as topological quantum computation\cite{nayakreview,kitaev2003}. 
The essence of this approach is the following.  The wavefunction $A_n$ where we store the quantum information can {\it only} be altered by braiding quasiparticles around each other.    Presuming the temperature is low enough that no stray quasielectron-quasihole pairs are created, the quantum information is thus protected from small amounts of noise.  Computations correspond to moving the quasiparticles around in particular ways to implement desired unitary transformations\cite{nayakreview,Hormozi1,Simula}.   This motion need not be performed precisely, as any motion that is topologically equivalent to the desired motion has the same effect (up to an abelian phase).   

\subsubsection{Calculating the Braiding Matrix}
\label{subsub:braiding}

An important question is how one actually calculates the matrix $U$ in Eq.~\ref{eq:Umatrix} given a particular braid.   As above we  suppose that for each given set of positions $\{ w_1, \ldots, w_M \}$  there is an orthonormal basis of wavefunctions $|
\Psi_n(w_1, \ldots, w_M; z_1, \ldots, z_N)\rangle$.    Here we remind the reader that the $z$'s are physical coordinates of electrons, and the wavefunction must be single valued in these $z$'s.   However, the $w's$ are simply parameters of the wavefunction, and the wavefunction need not be single valued in these coordinates --- there may be branch cuts.  If we move the $w$ coordinates around and then bring them back to their original points, we may return on a different Riemann sheet.   We should still have a set of $D$ wavefunctions spanning the same space.  However, if we return on a different  Riemann sheet, we might describe the space in a different basis.  A simple example of this would be if we move $w_1$ all the way around $w_2$ and we find that $\Psi_1$ and $\Psi_2$ have switched with each other due to branch cuts\footnote{We could have holomorphic forms like $\Psi_{\pm} = 1 \pm \sqrt{w_1 - w_2}$ which exchange when $w_1$ goes around $w_2$.}.   This effect is known as the {\it monodromy} of the wavefunction, and is represented by a matrix ${\cal M}$  --- this is the explicit change in the wavefunctions that you get by simply moving the $w$ coordinates around each other. 
 
However, in addition to the monodromy, there is a second contribution to the unitary braiding matrix $U$ in Eq.~\ref{eq:Umatrix}.   This contribution is a ``Berry matrix" term.    Let the positions $w_i$ move adiabatically as a function of some continuous parameter $\tau$ (which may or may not be time) so we have $w_i(\tau)$.   The $D$-dimensional unitary matrix is then given explicitly by  the so-called Berry matrix
\begin{eqnarray} \label{eq:UPath}
& & {\cal B} =  {\cal P} \exp\left[i \int d\tau  \left \langle \Psi_n | \frac{\partial}{\partial \tau} | \Psi_m  \right\rangle \right]  \\
 &
 \equiv& 1 + i \int d\tau  \left\langle \Psi_n | \frac{\partial}{\partial \tau} | \Psi_m  \right\rangle    - \int \!\! d\tau 
 \int^\tau \!\! d\tau'  \sum_q \left\langle \Psi_n | \frac{\partial}{\partial \tau} | \Psi_q  \right\rangle \left\langle \Psi_q | \frac{\partial}{\partial \tau'} | \Psi_m  \right\rangle    + \ldots \nonumber   
\end{eqnarray}
where $\cal P$ stands for path ordering of the exponential, the first few terms of which are expanded out in the second line.    

This type of evaluation of the Berry matrix is something one can imagine doing numerically\cite{tserkovnyak,BarabanThesis,baraban,wuestienne}.   However, since integrals are often hard to do numerically, a practical way of evaluating this Berry matrix is given  by dividing the path along $\tau$ into many small piece $p=1, \ldots, {\cal N}$ with $\cal N$ very large (and we have assumed the final set of positions $w_j$ is the same as the initial set of positions).  We then have\footnote{This discrete product can in fact be used to define the path-ordered integral in the limit that we have many steps along the path.}
$$
 {\cal B} = \prod_{p=1}^{{\cal N}-1}
  \left\langle \Psi_{a}
 (\tau_{p+1}) \rule{0pt}{12pt} \right|\left.  \rule{0pt}{12pt} 
 \Psi_{b}(\tau_p)
 \right\rangle 
$$
where each term in the product on the right is a matrix with subscript $a,b$ (ranging from 1 to the dimension $D$), and these matrices must be multiplied together in order to give the final matrix $\cal B$.    Note that although the positions $w_i(\tau)$ at the last step $\tau_{\cal N}$ are the same as the positions at the first step $\tau_1$ the wavefunctions $\Psi_n(\tau_1)$ and $\Psi_n(\tau_{\cal N})$ need not be the same as each other due to the monodromy (although the space spanned by all $n$ must be the same space). 

The unitary transformation $U$ in Eq.~\ref{eq:Umatrix} is given by the product of the monodromy and the Berry matrix
\begin{equation}
\label{eq:MB}
 U = {\cal M} {\cal B}
\end{equation}
The Berry matrix is the result of the adiabatic motion of the particles, whereas the monodromy simply accounts for the fact that when we get the particles back to their original position we may be working in a different basis from when we started. 

Note that we are allowed to redefine (rotate) our basis of states arbitrarily at any point along the path (we have only promised that at each point along the path we have an orthonormal basis).   Such a redefinition can change both the monodromy and the Berry matrix.  However, the final product of the two, the unitary matrix $U$ is independent of such redefinitions. 

Numerical calculations of braiding statistics had been done for
quasiparticles in abelian quantum Hall
states\cite{kjoensberg1999numerical,jeon2003fractional,ZaletelandMong},
and more recently for non-abelian quantum Hall states as
well\cite{tserkovnyak,baraban,BarabanThesis,wuestienne}.

\subsubsection{Holomorphic Orthonormal Basis}
\label{subsub:holo}

A great simplification occurs if one works with a basis which is both holomorphic and orthonormal.   In the CFT approach it is assumed that the so-called conformal block basis is exactly such a holomorphic, orthonormal basis\cite{read2009nonabelian,gurarienayak,bondersongurarie}.   We will return to this assumption, as it is an important one.   However, for now, let us simply assume we have a holomorphic orthonormal basis of the following form: 
\begin{eqnarray} 
 \label{eq:psiphi}
& &  \Psi_n(w_1, \ldots, w_M; z_1, \ldots z_N) = \phi_n(w_1, \ldots w_M; z_1, \ldots z_N)  \\ & &~~~~~~~~~~~~~~~~~~~~~~~~~~~~~~~~~~~~~ \times \exp\left[-\frac{1}{4 \ell^2} \left( \sum_{n=1}^N |z_n|^2   + \frac{e^*}{e} \sum_{n=1}^M |w_m|^2 \right)  \right]  \nonumber
\end{eqnarray}
such that  
\begin{equation}
\label{eq:orthonormality}
 \langle \Psi_n | \Psi_m \rangle = \delta_{nm.}
\end{equation}
where $e^*$ is the charge of the quasiparticle,  and we have reinstated the magnetic length $\ell = 1/\sqrt{e B}$ with $\hbar =1$.   The wavefunction $\phi_n$ is holomorphic in all variables, and must be single-valued in the $z$'s but may have branch cuts with respect to the $w$ coordinates so there will generally be a nontrivial monodromy from moving the quasiparticles.  

Now we calculate the Berry matrix in Eq.~\ref{eq:UPath} explicitly, assuming we have a holomorphic normalized wavefunction.  For simplicity, let us assume we are moving only a single quasiparticle at postion $w$ while keeping the other quasiparticles fixed.    We write
\begin{equation}
\label{eq:dtau}
 {\partial_\tau} = \frac{\partial w}{\partial \tau}  {\partial_w} + \frac{\partial \bar w}{\partial \tau}  {\partial_{\bar w}}
\end{equation}
Since $\phi_n$ (in Eq.~\ref{eq:psiphi}) is holomorphic, the only contribution  to $\partial_{\bar w}$ comes from the derivative acting on the gaussian factor and we get 
\begin{equation}
\label{eq:dbarw}
 \langle \Psi_n | \partial_{\bar w} | \Psi_m \rangle =  - \delta_{nm} \frac{e^*}{e} \frac{w}{4 \ell^2}
\end{equation}
In order to evaluate $\partial_w$, we note that
$ \partial_w \langle \Psi_n | \Psi_m \rangle =0 $ since the basis is orthonormal for all values of $w$.    We thus have
\begin{equation}
\label{eq:dw}
 \langle \Psi_n | \partial_{w} | \Psi_m \rangle =  -(\partial_w \langle \Psi_n |) |\Psi_m\rangle = \delta_{nm} \frac{e^*}{e} \frac{\bar w}{4 \ell^2}
\end{equation}
where again we have used that the wavefunction $\phi$ is  holomorphic (so $\phi^*$ is antiholomorphic).   We then have the Berry phase expression (inserting Eq.~\ref{eq:dtau} into Eq.~\ref{eq:UPath})
\begin{eqnarray*}
{\cal B} &=& {\cal P}  \exp\left[ i \oint d\tau  \left\langle \Psi_n | \frac{\partial}{\partial \tau} | \Psi_m  \right\rangle \right] \\
 &=& \delta_{nm} \exp\left[ i \frac{e^*}{4 e \ell^2 }\oint (\bar w dw - w d\bar w)  \right] =  \delta_{nm} \exp\left[i  \frac{A e^*}{\ell^2 e} \right]
\end{eqnarray*}
where $A$ is the area surrounded by the path of the particle (which we obtained using the complex version of Stokes theorem) and this Berry matrix is now completely diagonal. 
This phase is precisely the expected Aharonov-Bohm phase we should get for moving a charge $e^*$ around a total magnetic flux $A B$. 

Thus if we work in a holomorphic orthonormal basis, the Berry matrix is simply the Aharonov-Bohm phase.  The matrix part of the braiding matrix $U$ in Eq.~\ref{eq:Umatrix} is entirely from the monodromy contribution in Eq.~\ref{eq:MB}.   Thus we can determine the braiding properties of a wavefunction simply by examining the branch cut structure of the holomorphic orthonormal wavefunctions\cite{gurarienayak,read2009nonabelian}.   

Thus we see the value of working with a holomorphic orthonormal basis.  In the conformal field theory approach, it is assumed that the so-called ``conformal blocks" generate just such a basis.  Explicit forms of the conformal blocks have been worked out for the case of the Moore-Read state in Refs.~\refcite{moore1991nonabelions,nayakwilczek,ardonne2010chiral}.
We will discuss this issue further in section \ref{sec:CFT} below.   However, we comment here that one can use some of the methods of conformal field theory to try to prove this assumption of orthonormality.  While the orthonormality cannot typically be rigorously proven, it can often be reduced to an equivalent statement about screening of some effective plasma which can then be checked by numerics.\cite{Moore89,nayakwilczek,nayakreview,gurarienayak,
read2011hall,bondersongurarie,BLOK}

One might worry that perturbations to the wavefunction might change the braiding properties.  One particular concern that was raised, for example, is that Landau level mixing destroys the nice holomorphic properties of the wavefunctions --- and could change the braiding statistics (or ruin their topological nature altogether)\cite{SondhiLandauLevel}.   However, it has been proven that small perturbations cannot change the non-abelian part of the braiding --- although the abelian geometric phase {\it can} in fact be altered\cite{SimonMixingBraiding}.

\section{Hall Viscosity}
\label{sec:HallVisc}

The Berry matrix calculation, as in Eq.~\ref{eq:UPath} describes how the wavefunction varies as some parameter changes adiabatically.  In the above case, we are concerned with the change of the wavefunction as quasiparticles are moved around.   However, the idea is more general:  for any wavefunction which is a function of parameters, a Berry matrix (or Berry phase) can be defined.

A very useful case to study is a Berry phase associated with a change in geometry of a system for the ground state wavefunction.  Here we will consider a sheer deformation of the wavefunction on a torus and we will use the Berry connection to calculate a certain non-dissipative sheer viscosity of the system, known as the Hall viscosity\cite{avron1995viscosity,HoyosReview}. 

We consider a quantum Hall system on a torus\cite{Haldane1985periodic} geometry.     To describe a torus, we take a complex plane and define a parallelogram via the four points 0, 1, $\tau$  and  $1+\tau$ where $\tau = \tau_1 + i \tau_2$ with $\tau_i$ real.  Identifying opposite edges of this parallelogram, we get a torus. We thus have a wavefunction which is a function of two parameters $(\tau_1,\tau_2)$. 

When a normalized wavefunction varies as a function of two parameters, we can define a connection 
\begin{equation}
\label{eq:connection1}
A_\mu = i \langle \Psi | \partial_\mu  | \Psi \rangle
\end{equation}
and then define a curvature scalar\footnote{One may be concerned in Eq.~\ref{eq:connection1} that for fractional quantum Hall systems we should be keeping track of the multiple ground states on the torus.  However, the inner product in Eq.~\ref{eq:connection1} turns out to be diagonal in this additional index because the different ground states have different quantum numbers in the cases of interest.}
$$
 F = \epsilon_{\mu \nu} (\partial_\mu A_\nu - \partial_\nu A_\mu)
$$
  The Hall viscosity is then given by 
$$
 \eta_H = -\frac{4 \tau_2^2}{A} F
$$
with $A$ the area of the system.  This link between a Berry curvature and a viscosity was uncovered in the 1990s in the case of the integer quantum Hall effect\cite{avron1995viscosity}, and was extended more recently to the fractional case\cite{read2009nonabelian,read2011hall,bradlyn2012kubo,hoyos2012hall}.

The Hall viscosity is a transport coefficient from classical fluid and elasto-dynamics.   One defines a stress tensor
$
 \sigma_{ab} 
$
via the force $f_a$ 
$$
 \dot g_a = f_a = -\partial_b \sigma_{ab}
$$
with $g_a$ the momentum density.  Then this stress can be expanded 
\begin{equation}
\label{eq:strain}
  \sigma_{ab} = -\lambda_{abcd} u_{cd} - \eta_{abcd} \dot u_{cd} + \ldots
\end{equation}
where $u_{cd}$ is the symmetric local strain tensor
$$
 u_{cd} = \frac{1}{2} (\partial_c u_d + \partial_d u_c)
$$
with $u_c$ the local displacement.  In Eq.~\ref{eq:strain}, $\lambda$ is the elasticity tensor (in an isotropic fluid $\lambda \propto \delta_{ab}\delta_{cd}$), and $\eta$ is the viscosity.  In an isotropic system $\sigma$ is symmetric, and hence $\lambda$ and $\eta$ are symmetric under exchange of their first two indices or under exchange of their last two indices.   

Let us now focus on the viscosity term.  Note the viscosity gives a response to $\dot u_{cd}$, and this can be rewritten as 
$$
 \dot u_{cd} = \frac{1}{2} (\partial_c v_d  + \partial_d v_c)
$$
in terms of the velocity field $v$ (which we would expect to be the important quantity for a fluid).   We can then further decompose the viscosity into a symmetric and antisymmetric piece under exchange of first two with last two indices
$$
 \eta_{abcd} = \eta^S_{abcd} + \eta^A_{abcd}
$$
where $\eta^S_{abcd} = \eta^S_{cdab}$ and $\eta^A_{abcd} = -\eta^A_{cdab}$.     For an isotropic fluid we expect
$$
 \eta^A_{abcd} = \eta^H (\delta_{bc} \epsilon_{ad} - \delta_{ad} \epsilon_{bc}) 
$$
where we call $\eta^H$ the Hall viscosity\cite{avron1995viscosity,read2009nonabelian,read2011hall,hoyos2012hall,bradlyn2012kubo}.   Like the Hall conductivity, the antisymmetric part here is non-dissipative.   

Rather remarkably\cite{read2009nonabelian,read2011hall}, the Hall viscosity is not only related to the Berry phase but also turns out to be related to the intrinsic orbital angular momentum of the electrons.  The orbital angular momentum, in turn, is related to the shift $\cal S$ of the wavefunction (See Eq.~\ref{eq:shiftdef}), giving the general result
\begin{equation}
\label{eq:hallvis}
 \eta^H = \hbar \frac{\bar n {\cal S}}{4}
\end{equation}
where $\bar n$ is the average electron density. 

The Hall viscosity appears, not only in the fluid dynamics equations, but also as a coefficient of the electromagnetic response of the system in various limits.  For example, neglecting Zeeman energy we have the low frequency limit of the Hall conductivity at finite wavevector $q$ given by\cite{bradlyn2012kubo,hoyos2012hall,levinson}
$$
 \sigma_{xy}(q) = \frac{e^2}{\hbar} \left( \frac{\nu}{2 \pi} +  \left[\eta^H - 
\hbar \bar n  \right] (q \ell)^2 + \ldots  \right)
$$
where $\bar n$ is the particle density.

\section{Comments on Conformal Field Theory}
\label{sec:CFT}

Although much of the point of this review was to attempt to discuss recent developments without resorting to conformal field theory, it is probably worth making some comments on the CFT mapping.  Starting with the work of Moore and Read\cite{moore1991nonabelions},  this approach to understanding  quantum Hall states has been increasingly influential in the field.   While certain techniques related to CFT can be applied to almost any FQH state\cite{HanssonRMP,nayakreview}, we will see below that the states which are defined by special parent Hamiltonians fit the CFT framework much better than states that do not have such Hamiltonians.

\subsection{The Bulk-Edge Correspondence: In Brief}

A conformal field theory is generally defined as a quantum field theory that is invariant under conformal transformations\cite{Senechal97}.   In 1+1 dimensions (the only case that we will consider) this conformal invariance is such a strong restriction that many properties of such field theories can be obtained exactly.  It is crucial to know that 1+1 dimensional CFTs are very closely related to 2+1 dimensional topological quantum field theories\cite{Senechal97,nayakreview} --- a connection that is at the root of the Moore-Read approach.

The general idea of the FQHE-CFT connection is as follows: as a topologically ordered state of matter, one expects that much of the FQHE physics can
be described in terms of a topological quantum field theory --- a theory which is invariant under smooth deformations of space and
time\cite{nayakreview}.  To expose the interesting physics that results, we imagine examining a 2-dimensional slice of our $2+1$
dimensional space-time manifold.  Since the system is topologically invariant we expect that any direction we slice it will be equivalent.
What we mean by ``equivalent" here is that any type of slice should be
described by the same 2-dimensional quantum theory.  Let us label the three dimensions of our  space-time as $x,y$ and $t$ (for time).   If we slice the system at a fixed $y$-coordinate (giving us an
$x,t$ plane), the resulting theory is a $1+1$ dimensional CFT  which
describes the dynamics of the gapless FQHE edge.  On the hand if we slice the manifold at fixed time (giving us an $x,y$-plane) the resulting 2+0 dimensional static wavefunction is described by the {\it same} CFT  (we will explain below in more detail how we describe a wavefunction using a CFT).   This is an example (indeed,
it is perhaps the best explored example) of what is known as
``bulk-edge correspondence".  

While CFTs can be either unitary or nonunitary, only the unitary CFTs represent well behaved one dimensional dynamical systems.   (Nonunitary CFTs have, for example, scattering matrices that do not conserve amplitude).  Since the one dimensional edge can only be described by a unitary CFT, this implies that either the bulk-edge correspondence somehow fails, or the CFT must be unitary\cite{Read2009}.  One way in which this correspondence can indeed fail, is if the bulk is gapless so that edge excitations can leak into the bulk.  This gives the rough argument why gapped FQH states can only be described by unitary CFTs (although a number of other arguments have also been given\cite{read2009nonabelian}).  

We now explain a bit more detail of the FQHE-CFT correspondence as it is described in the original Moore-Read work\cite{moore1991nonabelions}.  In short, given an appropriate $1+1$ dimensional CFT we write a $2+0$ dimensional wavefunction as a correlator of the 1+1 dimensional theory
\begin{equation}
\label{eq:MR1}
 \Psi_{2d}(z_1, \ldots, z_N) = \langle \psi_e(z_1) \ldots \psi_e(z_N) \rangle_{1+1\, d}
\end{equation}
On the left, $z= x+  i y$ represents a position on a two dimensional plane, whereas on the right $z = x + i \tau$ is a space-time coordinate in a 1+1 dimensional chiral theory (the theory is chiral, depending only on $x+i \tau$, but not $x -  i \tau$, which when rotated back to real time gives a theory depending on $x - v t$ but not $x + v t$) with $v$ some velocity.   On the right  of Eq.~\ref{eq:MR1}, $\psi_e$ is an appropriately chosen operator within the 1+1 dimensional theory\footnote{For the experts we have not written the background charge operator for notational simplicity --- it is this background charge operator that gives the ubiquitous gaussian factors of the lowest Landau level wavefunctions, which are the only part of the wavefunction that is not holomorphic.  See the discussion in Ref.~\refcite{HanssonRMP} for example.}.  The details of this correspondence have been reviewed elsewhere\cite{nayakreview,HanssonRMP}.   Note that the electron field $\psi_e$ has a (conformal) scaling dimension (or conformal weight) called $h_e$ which describes the long and short distance limits of various correlation functions.   For example, we have\footnote{Also for the experts, here since we mean the $\psi_e$ field to contain both the charge and neutral sectors, the scaling dimension is the sum of charge and neutral scaling dimensions.}
$$
 \langle \psi_e(z) \psi_e^\dagger(z') \rangle = (z - z')^{-2 h_e}
$$
which (for a two point correlator) is correct at all distance.    This scaling dimension turns out to be related to the shift (see Eqs.~\ref{eq:shiftdef} and \ref{eq:hallvis}) of the quantum hall state (defined in Eq.~\ref{eq:MR1}) via the simple relation ${\cal S} = 2 h_e$.  We thus identify the scaling dimenson as being the mean orbital spin of the electron in the state (see the discussion near Eq.~\ref{eq:shiftdef}).

 The correspondence in Eq.~\ref{eq:MR1}, admittedly, looks a bit strange.  On the left, we have a wavefunction, whereas on the right, we have an expectation.    In trying to elucidate the meaning of this correspondence, it is perhaps useful to think about a slightly more general case.   Instead of thinking about a wavefunction on a disk geometry, let us think about an annulus (or similarly a cylinder)\cite{DubailReadRezayi}.  And instead of thinking about the ground state, let us think about a situation where both inner and outer edges of the annulus are in excited states.   Here we can write a more general wavefunction as
\begin{equation}
\label{eq:MR2}
 \Psi_{2d}^{u,v}(z_1, \ldots, z_N) = \langle u | \psi_e(z_1) \ldots \psi_e(z_N)  | v \rangle_{1+1\, d}
\end{equation}
where $u$ and $v$ are excited states of the 1+1 dimensional system.   This corresponds to an annulus with excitation $|u\rangle$ on the inner edge and excitation $|v\rangle$ on the outer edge.  We can think of these as the in-state and out-states of the 1+1 dimensional CFT.   

Returning now to a disk geometry (or an annulus where the inner edge remains in the ground state), we can re-examine Eq.~\ref{eq:edgeinner} and see that this is now rewritten as
\begin{eqnarray} \nonumber
 & &  \langle u | e^{-S} | v \rangle = \int d^2 {\bf z}_1  d^2 {\bf z}_2 \ldots  d^2 {\bf z}_N  \,\,  \langle u|  \psi_e^\dagger(z_N) \ldots \psi_e^\dagger(z_1) | 0 \rangle \, \langle 0 |  \psi_e(z_1) \ldots \psi_e(z_N) | v \rangle    \\
 &=& \nonumber \left\langle u \left|  
\int d^2 {\bf z}_1  d^2 {\bf z}_2 \ldots  d^2 {\bf z}_N  \psi_e^\dagger(z_N) \ldots \psi_e^\dagger(z_1) | 0 \rangle \, \langle 0 |  \psi_e(z_1) \ldots \psi_e(z_N)   
  \right| v \right\rangle,
\end{eqnarray}
the second line making the form of operator $e^{-S}$ obvious.  However, the fact that $S$ should be expandable in $1/N$ (and should vanish for large $N$) is not immediately obvious --- nor is the crucial statement that $S$ should be comprised of local operators from the edge CFT. 

Note that once we can write all edge excitations as excited states of the CFT, we can form coherent states of these excitations to make localized quasiholes.    This, however, can be rewritten using local operators from the CFT
\begin{equation}
 \Psi(w_1, \ldots, w_M; z_1, \ldots, z_N) = \langle \psi_{qh}(w_1) \ldots \psi_{qh}(w_M)  \psi_e(z_1) \ldots \psi_e(z_N) \rangle
\end{equation}
where $\psi_{qh}$ is an appropriately chosen local operator from the 1+1 dimensional CFT.   As discussed above in section \ref{subsub:braiding}, we expect that such an expression does not typically represent a single wavefunction but rather represents a vector space of conformal blocks.  

One can also now rephrase the orthormality hypothesis in Eq.~\ref{eq:orthonormality}.  Written out in terms of the 1+1 dimensional CFT, we are asking that for any fixed $w$'s, we have 
$$
 \delta_{nm} = \int d^2 {\bf z}_1 \ldots  d^2 {\bf z}_N  \,  \Psi^*_n(w_1, \ldots, w_M; z_1, \ldots, z_N)  \,  \Psi_m(w_1, \ldots, w_M; z_1, \ldots, z_N)
$$
where here the subscript $n$ and $m$ indicates which conformal block we are considering.      It is this statement that has been examined in depth in Refs.~\refcite{bondersongurarie,read2009nonabelian,BarabanThesis}.

\subsection{These Few CFTs are Apparently Special}
\label{sub:specialCFT}

One can try to build a quantum Hall wavefunction based on many different CFTs\cite{ESTIENNE2010} --- indeed, one only needs to find a CFT with an acceptable operator to act as an electron.   While this might sounds simple, in fact, it is not at all clear how many CFTs are actually physically acceptable.   With the exception of the Read-Rezayi series (including Laughlin and Moore-Read states) there remains no single-component gapped quantum Hall state generated by a CFT which has all of the nice properties we desire --- in particular having a special parent Hamiltonian such that all of the quasiholes of the system have zero interaction energy, and are described by the particle types of the CFT as well. 

While construction of such special Hamiltonians corresponding to CFTs has been shown possible for certain gapless states\cite{Gaffnian,BernevigHaldane1,dmitrygreenthesis}, and it looks potentially possible that such a construction might be possible for the (unitary) tricritical Ising model\cite{Jackson}, this has yet to be fully achieved, and it is still not clear that unitarity (and rationality) is sufficient to obtain a gapped quantum Hall state.  

Nonetheless, given the close relationship between CFTs and topological quantum field theories in general, it would not be surprising if such a construction could someow be achieved for any gapped fractional quantum Hall state.  Note that substantial progress towards such a construction for composite fermion states has been achieved recently\cite{HanssonRMP} although it falls somewhat short of what has been achieved for the Read-Rezayi series.  In particular, there is no special parent Hamiltonian, and the wavefunction is not simply written in the form of Eq.~\ref{eq:MR1}, but rather contains a symmetrization or antisymmetrization over a correlator of (at least two) different types of electron field.  These shortcomings make further analytic work on these wavefunctions extremely challenging.   While better CFT construction of these  states may not be possible, it remains tempting to try to find another route.   For example, the existence of a special parent Hamiltonian for an unprojected version of a hierarchy state\cite{JainEarly,RezayiMacdonald,Seidel25} seems a possibly promising approach. 

\subsection{Further Issues with the CFT Approach}

In trying to connect more closely to experiment, there are several further concerns related to the CFT approach.  For systems with multiple edge modes, a theory can only be conformally invariant if all of the edge modes move at the same velocity\footnote{Indeed, with very few conditions added (unitarity, Lorentz invariance, scale invairance) one can assume that any edge which has a single velocity can be described by a CFT.}.   While systems with special parent Hamiltonians and quadratic confinement (See section \ref{sec:edge}) necessarily have edge modes with the same velocity (which is determined only by the confining potential), more realistic systems generally can have multiple edge velocities.   However, as long as all the edge modes are moving in the same direction, this does not typically present a large problem  since one can at least imagine deforming the edge potentials locally so as to change the edge velocities arbitrarily.  In such cases many of the tools of CFT remain applicable even though the system lacks true conformal invariance.   

In more complicated cases edge modes move may in both directions. This occurs, for example, for the experimentally important Jain states with $\nu > 1/2$.     In this case we must treat both chiralities of the CFT, the right-movers, which are the holomorphic part and the left-movers, which are the anti-holomorphic part\footnote{In some cases one might be able to particle-hole conjugate a system to obtain a different state where all edge modes move in the same direction.  This is applicable, for example, for the Jain series at $\nu > 1/2$.  However, one can certainly find cases such as $\nu=2/7$ with edge modes moving in both directions, where this cannot be done.}.    This then becomes a bit more subtle since the bulk 2+0 dimensional wavefunction remains holomorphic, and the Moore-Read mapping (Eq.~\ref{eq:MR1}) obviously cannot hold precisely, although one might still assume that a topological connection between the edge and bulk theory should still hold.

\section{Conclusions}
\label{sec:conclusions}

This review has discussed a number of (mostly recent) advances in the understanding of certain fractional quantum Hall wavefunctions and their properties mainly by focusing on the explicit structure of the wavefunctions.  To a large extent we focused on wavefunctions that have ``special" properties of some sort --- usually meaning that they are zero energy states of a  special parent Hamiltonian.   Such wavefunctions have an enormous amount of mathematical structure that allows understanding at a deep and detailed level which would not be otherwise possible.   It is possible, however, that many other fractional quantum Hall states may show very similar properties if examined in the right light.  For example, we analytically understand an enormous amount about the Laughlin edge.    While we do not have the same analytic handle on general hierarchy or composite fermion states, for example, we expect that many of the same principles apply, and can at least be tested numerically.     Even more excitingly, many of the ideas that we can analytically explore in these quantum Hall states may extend to the study of other types of matter altogether.  For example, ideas of entanglement spectroscopy, which started in the quantum Hall field\cite{LiHaldane}, are now very widespread. 

The ideas presented in this review show many aspects of the deep mathematical structure present in fractional quantum Hall states.  The fact that the study of this structure has continued to develop for well over thirty years is a testament to its richness.  Indeed, despite the more than 160 citations included in this review, there are a certainly a huge number of additional closely related works that have not been discussed.  One should not interpret this omission as being an opinion that some of these excluded works are somehow less interesting, but rather that they simply did not fit within the particular narrative that I have chosen to follow.  Another author certainly could have made other choices and given different emphasis.  What is undoubtedly the case, no matter  which path  one choses to pursue through the field, is that this topic is full of interesting physics and interesting mathematics --- and it is very likely to continue to provide interesting directions for many years to come.

\bibliographystyle{ws-rv-van}
\bibliography{simonchapterrefs}


\end{document}